\documentclass[letterpaper,english,reprint, aps]{revtex4-1}
\usepackage[T1]{fontenc}
\usepackage[latin9]{luainputenc}
\usepackage{array}
\usepackage{multirow}
\usepackage{amsmath}
\usepackage{amssymb}
\usepackage{graphicx}
\usepackage{hyperref}

\makeatletter

\@ifundefined{pageheight}{\let\pageheight\pdfpageheight}{}
\@ifundefined{pagewidth}{\let\pagewidth\pdfpagewidth}{}
\pageheight\paperheight
\pagewidth\paperwidth

\providecommand{\tabularnewline}{\\}

\makeatother

\usepackage{babel}
\usepackage{color}

\newcommand{\COMMENTED}[1]{}

\begin{document}



\title{\textit{Ab initio} electronic density in solids by many-body \\ 
plane-wave auxiliary-field quantum Monte Carlo calculations}

\author{Siyuan Chen}
\email{schen24@email.wm.edu}

\selectlanguage{english}%


\author{Mario Motta}
\thanks{Present address: IBM Quantum, IBM Research Almaden, 650 Harry Rd, San Jose, CA 95120, USA}

\affiliation{Department of Physics, College of William \& Mary, Williamsburg,
Virginia 23185, USA}


\author{Fengjie Ma}

\affiliation{The Center for Advanced Quantum Studies and Department of Physics, Beijing Normal University, Beijing 100875, China}

\author{Shiwei Zhang}

\affiliation{Center for Computational Quantum Physics, Flatiron Institute, New
York, New York 10010, USA}

\affiliation{Department of Physics, College of William \& Mary, Williamsburg,
Virginia 23185, USA}
\begin{abstract}
We present accurate many-body results of the electronic densities in several 
solid materials, including Si, NaCl, and Cu. These results are obtained using the \textit{ab initio} auxiliary-field
quantum Monte Carlo (AFQMC) method working in a plane-wave basis with
norm-conserving, multiple-projector pseudopotentials. AFQMC has been shown to be an excellent
many-body total energy method. Computation of observables and correlation
functions other than the ground-state energy requires back-propagation,
whose adaption and implementation in the plane-wave basis AFQMC framework are discussed in the present paper. This
development allows us to compute correlation functions, electronic
densities and interatomic forces, paving the way for geometry optimizations
and calculations of thermodynamic properties in solids. Finite supercell size effects are considerably more subtle
in the many-body framework than in independent-electron calculations.
We analyze
the convergence of the electronic density, and obtain best estimates for 
the thermodynamic
limit.
The densities from several typical density functionals are benchmarked
against our near-exact results. The electronic densities we have obtained
can also be used to help construct improved density functionals. 

\COMMENTED{
\begin{description}
\item [{Usage}] Secondary publications and information retrieval purposes.{\small \par}
\item [{PACS~numbers}] May be entered using the environment \textsf{PACS~numbers}.{\small \par}
\item [{Structure}] You may use the \texttt{Description} environment to
structure your abstract.{\small \par}
\end{description}
}

\end{abstract}



\maketitle

\section{\label{sec:intro}Introduction}

The electronic density is one of the most fundamental physical quantities
in materials. 
Various structural properties in solids directly depend on the density and its closely related quantities.
More accurate results for the electronic density would thus lead to key improvement in our ability to reliably predict such physical properties.
Furthermore, the widely applied density functional theory (DFT)\citep{Hohenberg_PR_1964,Jones_RMP_2015,Becke_JCP_2014,Burke_JCP_2012}
relies on density functionals to yield an independent-electron 
approximation of the many-body
Hamiltonian. By iteratively solving the Schr\"{o}dinger equation using
such a density functional, the DFT method achieves high accuracy in
many systems, while simultaneously maintaining a relatively low computational
cost due to its one-body nature. As density functionals are designed
based on electronic 
densities, accurate density input is also essential for methodological development in DFT
\citep{Ceperley_PRL_1980,Stoudenmire_PRL_2012}.

Approaching an exact electronic density using correlated many-body
methods has remained a major challenge, despite the remarkable progress
witnessed in the last decades.
Exact methods like full-configuration interaction (FCI) require a
computational cost scaling exponentially with system size, and are
therefore usually restricted to systems of a small number of 
active electrons
and orbitals \citep{Vogiatzis_JCP_147}. A promising route for 
very accurate calculations is with quantum Monte Carlo (QMC) methods, which
represent the physical properties of a quantum system as multi-dimensional
integrals that are, in turn, evaluated using random sampling. 
To date the most widely used QMC
method for solid-state calculations has been the diffusion Monte Carlo
(DMC) approach \citep{Foulkes_RMP_2001,Reynolds_JCP_1982}, which reformulates
the imaginary-time Schr\"{o}dinger equation as a diffusion process by
a set of real-space configurations (walkers, given by electronic coordinates). The diffusion process
is guided by the kinetic energy, whereas the 
electron-electron interaction and the potential from nuclei are
represented by weights.

Auxiliary-field quantum Monte Carlo (AFQMC) is a more recent QMC approach for real materials \citep{Zhang_PRL_2003,Motta_WIRES_2018}.
AFQMC is based on the second-quantization scheme, and uses imaginary-time
evolution to propagate a trial wave function towards the desired ground
state. The auxiliary-field formalism has its roots in the Hubbard-Stratonovich
representation of the imaginary-time evolution \citep{Hubbard_PRL_1977},
which has long been applied in lattice field-theoretic calculations \citep{Sugiyama_ANN_1986,Sugar_PRD_1981,Sugar_PRB_1981}.
A reformulation to cast the projection as open-ended random walks in Slater determinant space \citep{Zhang_PRB_1997,Zhang_PRL_2003} and the conceptual connection with 
mean-field projection \citep{Zhang2018-Handbook} have provided the 
basis for coupling to standard electronic structure machinery, and positioned it 
as a natural post-DFT approach for  real materials. 
The AFQMC can incorporate pseudopotentials
in a simple manner 
that does not require any 
additional approximation \citep{Suewattana_PRB_2007}. Formally it allows 
computation of properties other than the total energy straightforwardly 
via back-propagation \citep{Zhang_PRB_1997,Purwanto_PRE_2004,Motta_JCTC_2017}.
As a low-polynomial scaling method, AFQMC has been shown in the
latest benchmarks 
to maintain excellent accuracy across a broad range of application areas \citep{LeBlanc_PRX_2015,Zheng_Science_2017,Motta_PRX_2017,Williams_PRX_2020,Shee_JCTC_2019}.

A set of orbitals (known as basis sets) are required
in AFQMC to express the Slater determinants. 
Similar to most other electronic structure methods, the basis sets could be plane
waves or local orbitals like Gaussians or Wannier functions \citep{AlSaidi_JCP_2006,Ma_PRL_2015}.
Since they can offer compact descriptions of electronic wave functions,
local orbitals are widely used in real materials and especially in
molecules. On the other hand, plane waves have several important
advantages that make them compelling, especially in crystalline solids:
(1) convergence to the infinite basis set limit requires only increasing
a single parameter, the kinetic energy cutoff; (2) plane waves are
orthonormal, which is a safeguard against numerical instabilities
seen for example in high-density systems \citep{Motta_PRX_2020}; 
(3) plane waves do not depend on
the positions of atoms in a cell and can be made to have very small finite basis error, hence Pulay corrections are absent
in force calculations \citep{Pulay_CompMolSci_2014}; 
(4) a large number of DFT calculations in extended systems are performed using
plane waves, and plane-wave AFQMC can use the same
computation framework as DFT [psuedopotentials, simple analytic evaluation  of matrix elements, fast Fourier transform (FFT), etc.].

Recent years of development and tests have indicated that plane-wave basis
AFQMC (PW-AFQMC) is an excellent total energy method \citep{Suewattana_PRB_2007,Purwanto_PRB_2009,Ma_PRB_2017}.
To calculate other physical quantities that do not commute
with the Hamiltonian, a back-propagation
technique is often necessary and needs to be incorporated in the algorithm \citep{Zhang_PRB_1997,Purwanto_PRE_2004,Motta_JCTC_2017}.
Among these quantities are the   electronic density and atomic forces, which have remained key challenges in extended systems of correlated electrons.

In this paper, we address this problem, by 
implementing back-propagation in PW-AFQMC and applying it to compute the electronic
densities in several crystalline solids: the widely studied covalent bond semiconductor 
silicon; 
the insulating ionic crystal sodium chloride (NaCl); and the transitional metal copper.
AFQMC calculations are carried out in different supercell sizes.
The computed densities are extrapolated to the  infinite supercell size or bulk limit using corrections from
a  finite-size DFT functional  \citep{Kwee_PRL_2008} and confirming numerical convergence with respect to the different supercell choices.
The final results 
 provide an accurate dataset of the true electronic densities for these systems. 
We compare these to the results  from 
several of the widely used  exchange-correlation functionals in DFT
to benchmark their
quality in reproducing the correct densities. 
We find that the accuracy of a functional in predicting the electronic density 
is not always correlated with its performance in computing total energies and structural properties. 

The rest of this paper is organized as follows. In Sec. \ref{sec:method},
we describe our method, including a brief overview of the AFQMC method,
the plane-wave basis implementation, and the state-of-the-art back-propagation
technique for observable computations. Section \ref{sec:comp-details}
introduces the systems we study 
and the computational details, 
including the finite-size correction and 
the details and the effect of the pseudopotential. 
Section \ref{sec:density}
presents the density results we have obtained, in diamond-structured
Si, NaCl, and fcc-Cu. 
In Sec. \ref{sec:benchmark} 
we describe our benchmark of several density functionals, by comparison in Sec.~\ref{subsec:density-benchmark} of their computed electronic densities with the AFQMC reference, followed by a discussion in Sec.~\ref{subsec:discussions} of the relation between the accuracy of a functional in computing the density versus the total energies.
We then conclude in Sec. \ref{sec:conclusion}.

\section{\label{sec:method}Methodology}

In this section we first provide a brief overview
of the plane-wave AFQMC method.
We specify the plane-wave Hamiltonian in detail, which  illustrates
the advantages of using a plane wave basis in solids. We then discuss
the phaseless AFQMC method, which directly leads to mixed-estimator
calculations for energy and properties.
This is followed by an introduction of the path-restoration back-propagation technique
formulated for plane-wave AFQMC, which is used to obtain higher-accuracy estimates of observables and correlation functions. 
This method is applied to obtain the charge densities.
\subsection{Plane-wave Hamiltonian}

Under the Born-Oppenheimer approximation \citep{Born_QThy_1927,Kittel_SSPhys_1996,Martin_ES_2020},
it is assumed that the ions move slowly compared to electrons.
It is then possible to approximate the solid-state Hamiltonian as a sum
of ionic and electronic parts,
\begin{equation}
H_{\mathrm{tot}}=V_{\mathrm{II}}+H\,,
\end{equation}
where $V_{\mathrm{II}}$ includes the ionic energies. 
Unless specified explicitly, we will assume Rydberg atomic units  throughout this paper,
$\hbar=2m_{e}=e^{2}/2=1$.
Using the plane-wave basis and under periodic boundary conditions,
the electronic Hamiltonian $H$ can be written as a sum of following components:
\begin{equation}
H=K+V_{\mathrm{psp}}+V_{\mathrm{Ewald}}+V_{\mathrm{Coulomb}}\,,
\end{equation}
%
where the first term is the kinetic energy, and the second
term $V_{\mathrm{psp}}$ is the nonlocal pseudopotential used to represent
the electron-ion interaction.
A kinetic energy cutoff $|\mathbf{G}|^{2}<E_{\mathrm{cut}}$ is imposed
on the plane waves, limiting the total number of plane waves to a
finite number $N_{\mathrm{PW}}$. 

The remaining terms describe the electron-electron interaction. The
Ewald term $V_{\mathrm{Ewald}}$ is from the self-interaction of a electron
with its periodic images, and is constant for a given lattice \citep{Kittel_SSPhys_1996,Martin_ES_2020,Ewald_AnnPhys_1921,DeLeeuw_RS_1980,DeLeeuw_RS_1980b}.
The operator
$V_{\mathrm{Coulomb}}$ is the electron-electron
Coulomb interaction and is given in the second-quantization as:
\begin{widetext}\[
V_{\mathrm{Coulomb}}=\frac{1}{\Omega}\sum_{ijkl}'\frac{4\pi}{|\mathbf{G}_{i}-\mathbf{G}_{k}|^{2}}\delta_{\mathbf{G}_{i}-\mathbf{G}_{k},\mathbf{G}_{l}-\mathbf{G}_{j}}\delta_{\sigma_{i},\sigma_{k}}\delta_{\sigma_{j},\sigma_{l}}c_{i}^{\dagger}c_{j}^{\dagger}c_{l}c_{k}\,,
\]\end{widetext}
where the prime in the sum indicates the exclusion of the $\mathbf{G}_{i}=\mathbf{G}_{k}$
singularity, $\sigma$ is the spin of the electron occupying the basis,
and italic letter subscripts such as 
$i$ represent the $(\mathbf{G}_{i},\sigma_{i})$
pair. This formula can be written as
\begin{equation}
V_{\mathrm{Coulomb}}=\frac{1}{\Omega}\sum_{\mathbf{Q\neq0}}\frac{4\pi}{|\mathbf{Q}|^{2}}\hat{\rho}^{\dagger}(\mathbf{Q})\hat{\rho}(\mathbf{Q})+V_{\mathrm{Coulomb,1b}}\,,\label{eq:vcoulomb}
\end{equation}
where $V_{\mathrm{Coulomb,1b}}$ is a one-body term from
$\mathbf{G}_{i}=\mathbf{G}_{l},\mathbf{G}_{j}=\mathbf{G}_{k}$. In
the first term, we introduced the one-body density operator $\hat{\rho}(\mathbf{Q})$,
which is defined as
\begin{equation}
\hat{\rho}(\mathbf{Q})=\sum_{\mathbf{G},\sigma}c_{\mathbf{G}+\mathbf{Q},\sigma}^{\dagger}c_{\mathbf{G},\sigma}\Theta(E_{\mathrm{cut}}-|\mathbf{G}+\mathbf{Q}|^{2})\,,\label{eq:density-op}
\end{equation}
where the step function ensures that $\mathbf{G}+\mathbf{Q}$ falls within the cutoff.
This leaves a finite number $N_{\mathrm{Q}}$ of distinct $\hat{\rho}(\mathbf{Q})$'s,
equal to the number of $\mathbf{Q}$-vectors ($\mathbf{Q}=\mathbf{G}_{1}-\mathbf{G}_{2}$).

\subsection{Auxiliary-field Quantum Monte Carlo (AFQMC)}

The basic idea underlying AFQMC is imaginary time propagation \citep{Zhang_PRB_1997}.
It is easy to prove that an initial state $|\phi\rangle$ which has
a nonzero overlap with the ground state |$\Psi_{0}\rangle$ propagates
to the ground state with the following projection:
\begin{equation}
|\Psi_{0}\rangle\sim\lim_{\beta\to\infty}e^{-\beta H}|\phi\rangle=\lim_{N\to\infty}e^{-N\Delta\tau H}|\phi\rangle\,,
\end{equation}
where we separate the total propagation time $\beta$ into
$N$ time-steps of 
$\Delta\tau$, and perform the projection
iteratively until convergence is reached. In fermion systems, usual
choices of the wave functions are single or multiple Slater determinants.
To turn this imaginary time evolution into a practical algorithm,
we need to implement a single-step propagation $e^{-\Delta\tau H}$.
For small enough $\Delta\tau$, we can split up the components in
the Hamiltonian using the Trotter-Suzuki decomposition:
\begin{equation}
e^{-\Delta\tau H}=e^{-\frac{\Delta\tau}{2}H_{1}}e^{-\Delta\tau H_{2}}e^{-\frac{\Delta\tau}{2}H_{1}}+O(\Delta\tau^{3})\,,\label{eq:Trotter}
\end{equation}
%
where $H_{1}$ and $H_{2}$ denote the one-body and two-body part
of the Hamiltonian, respectively. The one-body propagation is a simple
and computationally inexpensive operation to perform, as it maps a
Slater determinant to another Slater determinant, according to Thouless'
theorem \citep{Thouless_NuclPhys_1960,Thouless_NuclPhys_1961}. To
propagate in imaginary-time under $H_{2}$, we will make use of the particular structure in the plane-wave basis, 
in Eq. (\ref{eq:vcoulomb}):
\begin{equation}
H_{2}=\frac{1}{\Omega}\sum_{\mathbf{Q\neq0}}\frac{4\pi}{|\mathbf{Q}|^{2}}\hat{\rho}^{\dagger}(\mathbf{Q})\hat{\rho}(\mathbf{Q})\,.
\end{equation}

Using the identity $\hat{\rho}(\mathbf{Q})=\hat{\rho}^{\dagger}(-\mathbf{Q})$,
we can write it into a sum of squares of one-body operators:
\begin{equation}
\label{eq:H2-sumsq}
H_{2}=\frac{1}{4}\sum_{\mathbf{Q\neq0}}[\hat{\alpha}^{2}(\mathbf{Q})+\hat{\beta}^{2}(\mathbf{Q})]\,,
\end{equation}
where the one-body Hermitian operators $\hat{\alpha}(\mathbf{Q}),\hat{\beta}(\mathbf{Q})$
are linear combinations of $\hat{\rho}(\mathbf{Q})$ and $\hat{\rho}^{\dagger}(\mathbf{Q})$ \cite{Suewattana_PRB_2007}.
We then perform a continous-variable Hubbard-Stratonovich transformation:
\begin{equation}
e^{-\frac{\Delta\tau}{2}\lambda\hat{v}^{2}}=\frac{1}{\sqrt{2\pi}}\int_{-\infty}^{\infty}dxe^{-\frac{1}{2}x^{2}}e^{x\sqrt{-\Delta\tau\lambda}\hat{v}}\,,\label{eq:HStrans}
\end{equation}
where $\hat{v}$ is $\hat{\alpha}(\mathbf{Q})$ or $\hat{\beta}(\mathbf{Q})$,
and $\lambda$ is a constant. 
This 
transforms the
two-body 
propagator into an integral of one-body operators.
It represents $e^{-\frac{\Delta\tau}{2}\lambda\hat{v}^{2}}$
as an expectation value over a random variable $x$, called an {\em auxiliary
field}, with Gaussian probability distribution.
For the operator $H_{2}$ in Eq.~(\ref{eq:H2-sumsq}), $\mathbf{x}=\{x_{\mathbf{Q}}^{(\alpha)},x_{\mathbf{Q}}^{(\beta)}\}$
is a set of auxiliary fields corresponding to the one-body operators,
$\hat{\alpha}(\mathbf{Q})$ and $\hat{\beta}(\mathbf{Q})$, which commute and can be broken up and recombined in the exponential.
(However, the behavior of the Trotter error in AFQMC in the presence of constraints is more subtle,  as discussed in, e.g.,  
Refs.~\citep{Purwanto_PRB_2009,Shi_JCP_2021}.)
 The expectation
value in Eq.~(\ref{eq:HStrans}) can be evaluated by Monte Carlo techniques.

A simple scheme to propagate an initial state in imaginary time consists 
of setting 
a large number of Slater determinants $|\Phi_{0}^{(i)}\rangle$,
called walkers, to the initial wave function $|\phi\rangle$. For
each walker $i$ and imaginary-time step $k$, an auxiliary field
configuration $\mathbf{x}_{k}^{(i)}$ is sampled and the walker is
projected in imaginary time as $|\Phi_{k+1}^{(i)}\rangle=\hat{B}(\mathbf{x}_{k}^{(i)})|\Phi_{k}^{(i)}\rangle$,
where $\hat{B}(\mathbf{x})$ denotes the product of one-body propagators \cite{Zhang_PRL_2003}.

This simple algorithm, however, suffers from several 
problems
which we need to tackle individually.
A uniform random sampling
of the entire Hilbert space results in large statistical noise,
and needs to be replaced by importance sampling \citep{Zhang_PRL_2003,Zhang_Lecture_2013,Motta_WIRES_2018}.
AFQMC uses a trial wavefunction $|\Psi_{\mathrm{T}}\rangle$ to guide
the stochastic sampling of walkers, and attaches a weight $w_{k}^{(i)}$
to each walker which is designed to be proportional to the overlap
of the determinant $|\Phi_{k}^{(i)}\rangle$ with $|\Psi_{\mathrm{T}}\rangle$.
At each time step, the wave function is thus stochastically represented
in the following form:

\begin{equation}
|\Phi_{k}\rangle\sim\sum_{i}w_{k}^{(i)}\frac{|\Phi_{k}^{(i)}\rangle}{\langle\Psi_{\mathrm{T}}|\Phi_{k}^{(i)}\rangle}\,.
\label{eq:wf-imp-sampl}
\end{equation}

Walkers whose overlaps with $|\Psi_{\mathrm{T}}\rangle$ have large magnitudes are
deemed more important and are sampled more frequently, improving the
efficiency. Note that the ``importance function'' defined above is complex; inclusion of the phase information is crucial for the 
accuracy of the phaseless approach \citep{Purwanto_PRE_2004,Zhang_PRL_2003}.  
To realize the Monte Carlo sampling given in Eq.~(\ref{eq:wf-imp-sampl}), a modified probability is used by introducing a shift $\bar{\mathbf{x}}$ in the auxiliary field,
called {\em force bias}.
The optimal choice of the force bias is the mixed estimator of the one-body operators, 
$\hat{\alpha}(\mathbf{Q})$ and $\hat{\beta}(\mathbf{Q})$ \citep{Purwanto_PRE_2004}.
A local-energy is simultaneously introduced
$E_{L}(\Psi)=\langle\Psi_{\mathrm{T}}|H|\Psi\rangle/\langle\Psi_{\mathrm{T}}|\Psi\rangle$.
The final form of the weight factor is given by:
\begin{equation}
w_{n}^{(i)}=\prod_{k=0}^{n-1}I(\mathbf{x}_{k}^{(i)}-\bar{\mathbf{x}}_{k}^{(i)},\Phi_{k}^{(i)})\,,
\end{equation}
with
\begin{equation}
I(\mathbf{x}_{k}^{(i)}-\bar{\mathbf{x}}_{k}^{(i)},\Phi_{k}^{(i)})\simeq e^{\Delta\tau(E_{\mathrm{T}}-[E_{L}]_{k}^{(i)})}\,,\label{eq:weight-fac}
\end{equation}
where $E_{\mathrm{T}}$ is a trial energy chosen to approximate the exact energy and can be improved iteratively during the calculation. 
Below we will sometimes use $[E_{L}]_{k}^{(i)}$ as a shorthand for the local energy $E_L(\Psi_{k}^{(i)})$  
at step $k$  for walker $i$. 

The formulation of the method up to this point is exact, but 
suffers from a phase problem, as weights of walkers are eventually
randomly distributed into the entire complex plane and statistical
noise would increase exponentially with the number of time steps and the system size. 
Propagation of the random walkers in the branching random walk framework, with no additional intervention, is referred to as a \textit{free projection}. 
To control the phase problem, the solution is to make a \textit{phaseless approximation}
which introduces a small systematic bias, but reduces statistical
fluctuations from exponential to polynomial. The method can be formally represented by rewriting Eq.~(\ref{eq:weight-fac}) as
\begin{equation}
I(\mathbf{x}_{k}^{(i)}-\bar{\mathbf{x}}_{k}^{(i)},\Phi_{k}^{(i)})\simeq e^{\Delta\tau(E_{\mathrm{T}}-\mathrm{Re}[E_{L}]_{k}^{(i)})}\times\max\{0,\cos[\Delta\theta]_{k}^{(i)}\}\label{eq:weight-fac-phl}
\end{equation}
with
\begin{equation}
[\Delta\theta]_{k}^{(i)} \equiv \mathrm{Arg}\frac{\langle\Psi_{\mathrm{T}}|\hat{B}(\mathbf{x}-\bar{\mathbf{x}})|\Phi_{k}^{(i)}\rangle}{\langle\Psi_{\mathrm{T}}|\Phi_{k}^{(i)}\rangle}\,.\label{eq:phl-delta-th}
\end{equation}

Equation (\ref{eq:weight-fac-phl}) turns the weights into real and positive
quantities, 
and contains their
statistical fluctuations with the 
cosine factor. 
Walkers that are extremely close to the origin $\langle\Psi_{\mathrm{T}}|\Phi_{k}^{(i)}\rangle=0$
are the major contributors to the phase problem. These walkers move
rapidly around the origin, resulting in larger $\Delta\theta$  in Eq.~(\ref{eq:phl-delta-th}) and 
therefore 
a smaller cosine factor; this factor is zero if
$\theta>\frac{\pi}{2}$, meaning the walker is eliminated.

\subsection{Back-propagation and computation of charge density}

The quantum mechanical expectation value of an observable $\hat{O}$
over a state $|\Psi_{0}\rangle$ is
\begin{equation}
\langle\hat{O}\rangle=\frac{\langle\Psi_{0}|\hat{O}|\Psi_{0}\rangle}{\langle\Psi_{0}|\Psi_{0}\rangle}\,.
\end{equation}
Since the ground state is an eigenfunction of the Hamiltonian, $H|\Psi_{0}\rangle=E|\Psi_{0}\rangle$,
for $\hat{O}=\hat{H}$ the exact estimator coincides with the 
mixed estimator:
\begin{equation}
E=\langle H\rangle=\frac{\langle\Psi_{\mathrm{T}}|H|\Psi_{0}\rangle}{\langle\Psi_{\mathrm{T}}|\Psi_{0}\rangle}\,,
\end{equation}
where only the ket state is propagated in imaginary time.
In an AFQMC calculation, the energy at step $k$  is thus estimated  as
\begin{equation}
E=\frac{\sum w_{k}^{(i)} E_L(\Phi_{k}^{(i)})}{\sum_{i}w_{k}^{(i)}}\,.
\end{equation}

For observables not commuting with the Hamiltonian, mixed estimators
are biased, so the bra state has to be propagated in imaginary time
as well. To achieve this goal, Zhang and coworkers \citep{Zhang_PRB_1997,Purwanto_PRE_2004,Motta_JCTC_2017}
proposed a back-propagation (BP) method that rewrites the estimator
as
\begin{equation}
\langle O\rangle\simeq\frac{\langle\Psi_{\mathrm{T}}|e^{-m\Delta\tau H}\hat{O}e^{-n\Delta\tau H}|\Phi_{0}\rangle}{\langle\Psi_{\mathrm{T}}|e^{-(m+n)\Delta\tau H}|\Phi_{0}\rangle}\,,
\end{equation}
where $n$ is the number of \textquotedblleft forward propagation\textquotedblright{}
steps and $m$ is the number of \textquotedblleft back-propagation\textquotedblright{}
steps. The BP estimator reduces to the mixed estimator for $m=0$,
and gives higher-accuracy results for $m$ sufficiently large. The
backwards projection is performed on $|\Psi_{n,0}\rangle=|\Psi_{\mathrm{T}}\rangle$
by applying adjoints of propagators in reverse order,
\begin{equation}
|\Psi_{n,m}^{(i)}\rangle=\hat{B}^{\dagger}(\mathbf{x}-\bar{\mathbf{x}})_{n}^{(i)}\hat{B}^{\dagger}(\mathbf{x}-\bar{\mathbf{x}})_{n+1}^{(i)}...\hat{B}^{\dagger}(\mathbf{x}-\bar{\mathbf{x}})_{n+m-1}^{(i)}|\Psi_{\mathrm{T}}\rangle\,.\label{eq:bp-prop}
\end{equation}
Note that we only need to store the sampled auxiliary-fields on the path in order to recover the propagators.
The BP estimate of an observable is now given by
\begin{equation}
\langle O\rangle\simeq\frac{1}{\sum_{i}w_{n+m}^{(i)}}\sum_{i}w_{n+m}^{(i)}\frac{\langle\Psi_{n,m}^{(i)}|\hat{O}|\Phi_{n}^{(i)}\rangle}{\langle\Psi_{n,m}^{(i)}|\Phi_{n}^{(i)}\rangle}\,,\label{eq:bp-est}
\end{equation}
where the weights are evaluated at step $(n+m)$. In early
AFQMC back-propagation practices, the forward- and back-propagation
use the same technique, e.g. the \textit{free-projection BP}~involves
use of free projection on both sides, while the \textit{phaseless
BP} applies phaseless approximation to both directions 
in the imaginary time. 
Phaseless BP stabilizes the algorithm and prevents the onset of the phase problem,
but also introduces an unavoidable bias in the BP estimator. The phaseless
bias is more severe for back-propagation than forward propagation,
since the phaseless approximation breaks the symmetry between the
two propagation directions 
and is optimal only for the forward direction, to which its imaginary-time dependence is aligned.

To mitigate the phaseless bias in back-propagation, a technique called \textit{path-restoration}
was recently proposed \citep{Motta_JCTC_2017}. In the path-restoration
technique, for each time the auxiliary field and force bias are recorded,
we also record the discarded $\mathrm{Im}[E_{L}]$ and the cosine
factor $\cos(\Delta\theta)$ in Eq. (\ref{eq:weight-fac-phl}). If
this walker survives from step $n$ to step $n+m$ (i.e. its weight
remains non-zero), the information discarded due to the phaseless
approximation can be restored when computing observables in Eq.~(\ref{eq:bp-est}),  by multiplying the walker\textquoteright s
weight with a restoring factor:
\begin{equation}
\omega_{n+m}^{(i)}\rightarrow \omega_{n+m}^{(i)}\, f_{n':n+m}^{(i)}\,,
\label{eq:bp-pres-fac}
\end{equation}
where the path restoration factor is
\begin{equation}
f_{n':n+m}^{(i)} \equiv \prod_{k=n'}^{n+m}e^{-\Delta\tau\mathrm{Im}[E_{L}]_{k}^{(i)}}\frac{1}{\cos([\Delta\theta]_{k}^{(i)})}\,.
\end{equation}
The imaginary-time index $n'$ in Eq.~(\ref{eq:bp-pres-fac}) is typically chosen as $n'=n$, which corresponds to path-restoration 
only on the back-propagation portion of the path \citep{Motta_JCTC_2017}. We can also choose $n'>n$, which gives a partial restoration, if it is too 
noisy to restore the entire BP path. Conversely, it is possible to choose $n'<n$, which amounts to a partial restoration of the 
ket $|\Phi_{n}^{(i)}\rangle$ in Eq.~(\ref{eq:bp-est}). In other words, since $ f_{n':n+m}^{(i)}= f_{n:n+m}^{(i)} f_{n':n}^{(i)}$, the factor 
$ f_{n':n}^{(i)}$ can be grouped with $\omega_{n}^{(i)}$ to form a partial restoration  in the \textit{forward direction} of the path of length 
$(n-n')$ leading up to the ket. 
The path restoration produces weights closer to the free-projection form Eq. (\ref{eq:weight-fac}).
Due to this information recovery, it was found in several molecular systems
\citep{Motta_JCTC_2017} that path restoration BP provides more accurate
estimates of observables than phaseless BP, though leading to larger
statistical fluctuations. (Several of the results in Ref.~\citep{Motta_JCTC_2017} were actually 
obtained with $n'<n$ but mislabeled as $n'=n$.
\citep{Motta_private_communication})

The starting point to evaluate any one-body property is the
one-body density matrix, also called an equal-time Green\textquoteright s function,
\begin{equation}
{\mathcal G}_{ij}\equiv\langle c_{i}^{\dagger}c_{j}\rangle=\frac{\langle\Psi_{0}|c_{i}^{\dagger}c_{j}|\Psi_{0}\rangle}{\langle\Psi_{0}|\Psi_{0}\rangle}\,,
\end{equation}
where recall $i$ represents the $(\mathbf{G}_{i},\sigma_{i})$
pair for plane-wave basis functions. In fact, given the Green\textquoteright s
function, estimating any one-body observables (e.g. charge density
and forces) only requires a simple post-processing operation,
\begin{equation}
\langle\hat{O}_{\mathrm{1b}}\rangle=\sum_{j,k}(O_{\mathrm{1b}})_{jk}{\mathcal G}_{kj}=\mathrm{Tr}[O_{\mathrm{1b}}{\mathcal G}]\,.
\end{equation}

To evaluate the charge density within second quantization, we represent
the field operator as
\begin{equation}
\hat{\varphi}_{\sigma}^{\dagger}(\mathbf{x})=\sum_{\mathbf{G}}\varphi_{\mathbf{G}}(\mathbf{x})c_{\mathbf{G},\sigma}^{\dagger},\,\,\,\,\varphi_{\mathbf{G}}(\mathbf{x})=e^{i\mathbf{G}\cdot\mathbf{x}}\,,\label{eq:pw-field-op}
\end{equation}
and obtain the following expression for the charge density:
\begin{equation}
\rho(\mathbf{x})=\frac{\langle\Psi_{0}|\hat{\varphi}_{\sigma}^{\dagger}(\mathbf{x})\hat{\varphi}_{\sigma}(\mathbf{x})|\Psi_{0}\rangle}{\langle\Psi_{0}|\Psi_{0}\rangle}\,.
\end{equation}
Now, switching to plane wave basis using Eq.~(\ref{eq:pw-field-op}),
we obtain
\begin{equation}
\rho(\mathbf{x})=\sum_{ij}e^{i(\mathbf{G}_{j}-\mathbf{G}_{i})\cdot\mathbf{x}}\delta_{\sigma_{i},\sigma_{j}}\langle c_{i}^{\dagger}c_{j}\rangle\,.
\end{equation}
This allows us to relate the charge density at $\mathbf{x}$
to the Green's functions. In practice, we group elements of the Green's
function with the same $\mathbf{Q}=\mathbf{G}_{j}-\mathbf{G}_{i}$,
i.e. defining
\begin{equation}
\langle\hat{\rho}(\mathbf{Q})\rangle=\sum_{ij}\delta_{\mathbf{Q},\mathbf{G}_{j}-\mathbf{G}_{i}}{\mathcal G}_{ij}\,,
\end{equation}
which is
just an estimator of the one-body density
operator $\hat{\rho}(\mathbf{Q})$ in Eq.~(\ref{eq:density-op}). 
The reciprocal-space density operator is related to the 
real space charge density $\rho(\mathbf{x})$ 
by a Fourier transform,
\begin{equation}
\rho(\mathbf{x})=\sum_{\mathbf{Q}}e^{i\mathbf{Q}\cdot\mathbf{x}}\langle\hat{\rho}(\mathbf{Q})\rangle\,.
\end{equation}

The density can be computed by storing only the  density operators
$\hat{\rho}(\mathbf{Q})$, requiring memory $\mathcal{O}(N_{\mathrm{Q}})\approx\mathcal{O}(8N_{\mathrm{PW}})$.
Similarly, other one-body observables can be computed based on the Green's function. 
The memory requirement to store the Green's function will be no larger than the above for
a system with translational symmetry as our systems are. 

\section{\label{sec:comp-details}Computational Details}

In this section we describe several computational details of
our calculations. The first one is the 
correction used
to remove finite-size effect in the AFQMC density, extrapolating the
result to the thermodynamic limit. The second is the pseudopotential, 
which is part of the definition of our many-body Hamiltonian that affects the precise values of 
the electronic density we provide. 
In the last subsection, we include 
any additional details on
the three solid systems investigated.

\subsection{\label{subsec:FSCorr}Finite-size correction}

Finite-size
effects must be properly reduced and removed in many-body electronic
structure calculations of extended systems. 
Convergence to the thermodynamic limit is often slow 
while computational cost tends to grow quickly with respect to supercell size.
Common methods to address this problem
include performing an extrapolation using several different cell size,
or designing a finite-size correction method to reduce the finite-size
effect.

To simulate the finite-size effects in many-body calculations,
Kwee \textit{et al}.~\citep{Kwee_PRL_2008,Ma_PRB_2011} proposed a finite-size correction
method to the total energy by designing a finite-size LDA ($\mathrm{LDA^{FS}}$,
also called KZK) exchange-correlation functional. Parameters in the
functional were fitted using total energy results of the electron gas obtained in
finite cubic supercells of volume $V_{0}$. 
A correction to the energy per 
formula unit
can then be applied to obtain an improved estimate of the 
value at the thermodynamic limit:
\begin{equation}
E[\infty]\approx E^{\mathrm{QMC}}[V_{0}]-\{E^{\mathrm{LDA^{FS}}}[V_{0}]-E^{\mathrm{LDA}}[\infty]\}\,,
\end{equation}
where $E^{\mathrm{QMC}}[V_{0}]$ is the QMC result for the supercell with volume $V_0$ and at a single $k$-point or 
a set of $k$-points, $E^{\mathrm{\mathrm{LDA^{FS}}}}[V_{0}]$ is the corresponding calculation with the same 
supercell and $k$-point(s), using the finite-size LDA functional, and $E^{\mathrm{LDA}}[\infty]$ is from a standard LDA 
calculation using a converged dense $k$-point grid.
This finite-size correction method was found to deliver good 
finite-size corrections on various solid systems and allow quick convergence to the thermodynamic limit.

\begin{figure}[b]
\includegraphics[width=1\linewidth]{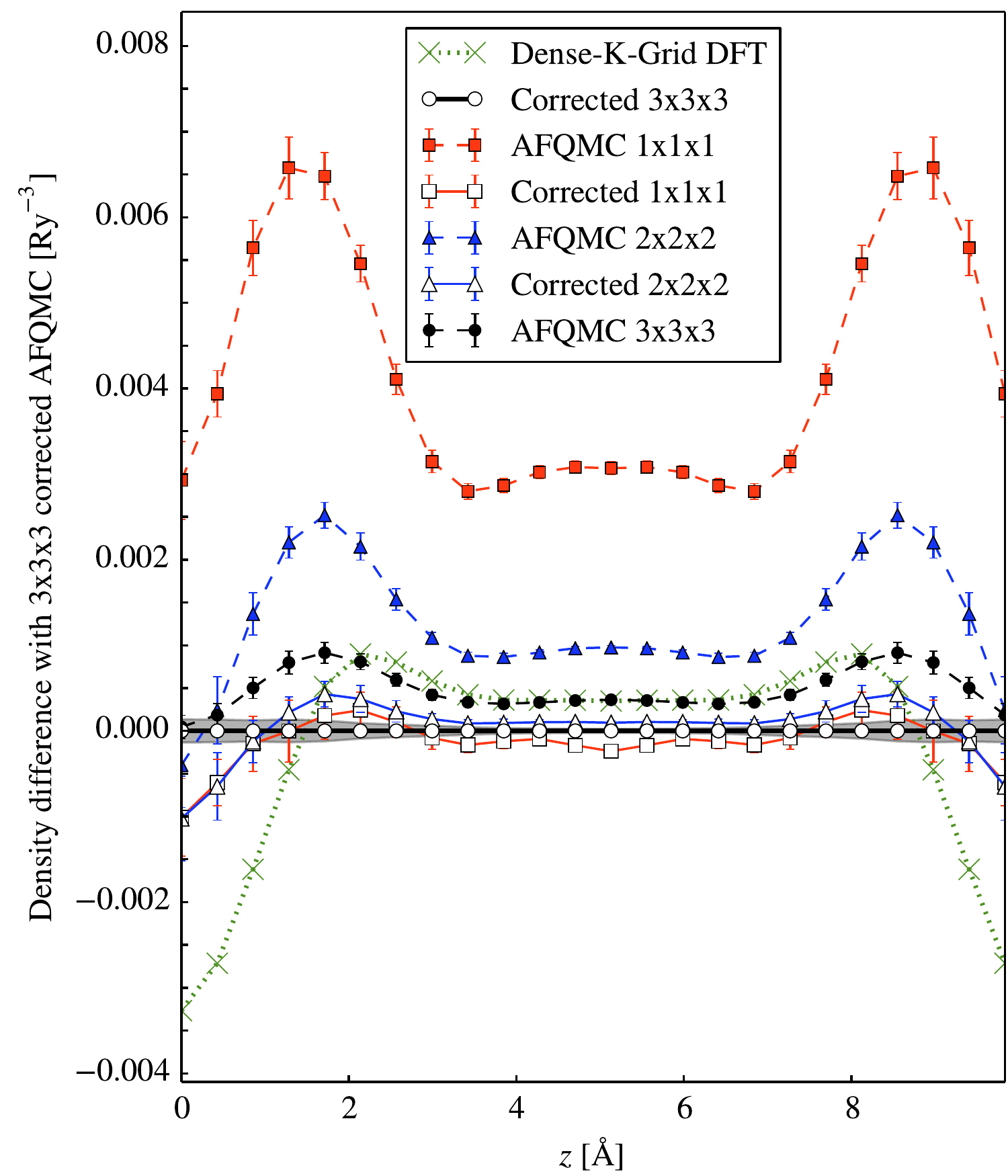}\caption{\label{fig:Si-FSCorr} Demonstration of the finite-size correction of the computed electronic density,
along a line cut in Silicon.
Deviations from the final result at the
thermodynamic limit are shown. Red, blue and black solid lines are AFQMC densities for supercell sizes $1\times1\times1$,
$2\times2\times2$ and $3\times3\times3$, respectively. Dashed lines
are uncorrected, ``raw'' AFQMC density, and the solid
lines are finite-size corrected AFQMC density. 
Note the error for $3\times3\times3$ corrected AFQMC density is plotted in shades.
For reference, the dotted green line is the dense-$k$-point ($6\times6\times6$)
grid DFT-LDA density. }
\end{figure}

In our work, we adapt the KZK finite-size correction technique, 
and extend the concept from total energy $E$
to charge density $\rho(\mathbf{x})$, defined for $\mathbf{x}$ within a formula cell:
\begin{equation}
\rho[\infty](\mathbf{x})\approx\rho^{\mathrm{QMC}}[V_{0}](\mathbf{x})-\{\rho^{\mathrm{\mathrm{LDA^{FS}}}}[V_{0}](\mathbf{x})-\rho^{\mathrm{LDA}}[\infty](\mathbf{x})\}\label{eq:fscorr}
\end{equation}
Using this method, we have observed accelerated convergence in all solid systems
we investigated. For example, Fig.~\ref{fig:Si-FSCorr} 
shows the effect of correction along a line cut in diamond-structured silicon
(see Sec.~\ref{subsec:systems} for details of the
system). Convergence of the AFQMC charge density with increasing cell
size is evident. Finite-size correction accelerates this convergence:
the corrected AFQMC charge density of a $1\times1\times1$ is almost
as close to the thermodynamic limit as the uncorrected  $3\times3\times3$ AFQMC result, and the corrected results from  $1\times1\times1$,  $2\times2\times2$,
and  $3\times3\times3$ supercells are essentially in agreement within statistical error.

Metals like Cu have particularly strong finite-size errors, which
result in non-negligible deviations between densities at different
$k$-points even after the finite-size correction 
is applied. To further reduce these residual finite-size errors, we
choose to perform a $k$-point averaging [i.e., using a set of $k$-points in Eq.~(\ref{eq:fscorr}) rather than a single one]. 
Quasi-random $k$-point sequences
are utilized to reduce the need of DFT smearing and increase the convergence
speed \cite{Qin_PRB_2016}. More computational details are listed in Section \ref{subsec:systems}.

\subsection{\label{subsec:psp-eff}Pseudopotential Core Effect}

\begin{figure}
\includegraphics[width=1\linewidth]{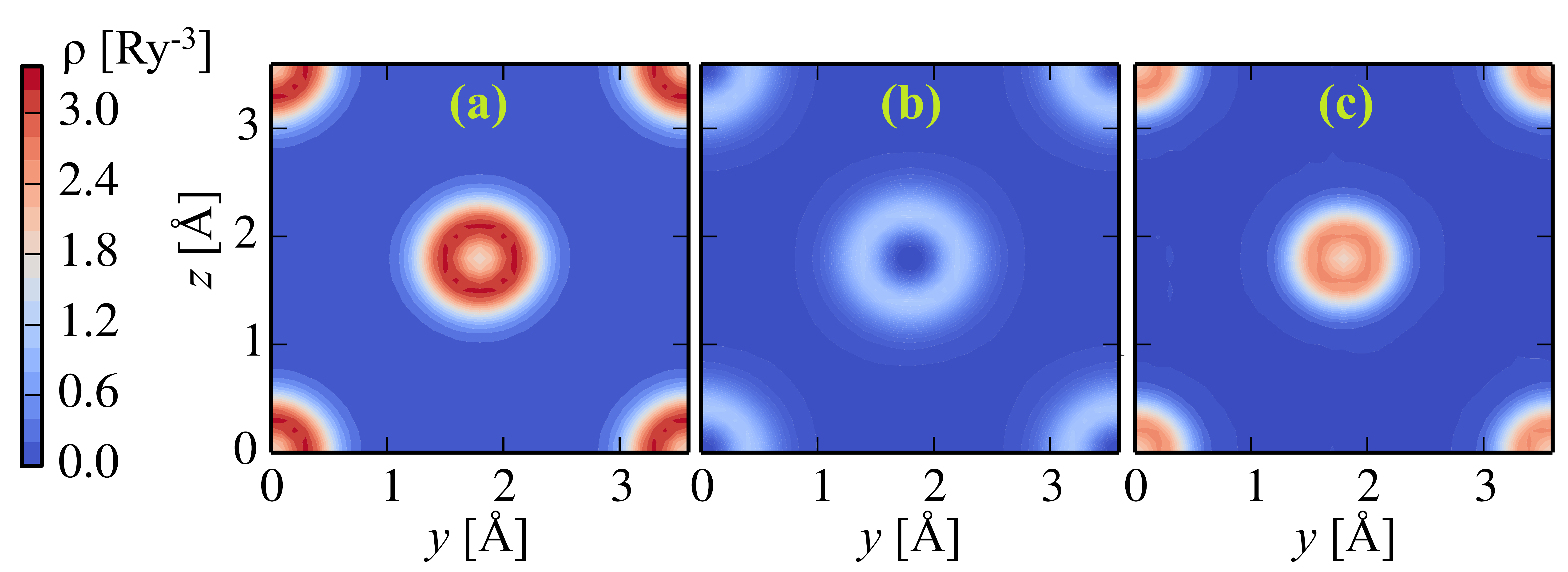}\caption{\label{fig:Cu-psp-eff}Pseudopotential effect in Cu: 
(a) PW-AFQMC density using a Ne-core pseudopotential (19 valence electrons);
(b) PW-AFQMC density using a Ar-core pseudopotential (11 valence electrons);
(c) difference between (a) and (b), showing
the contribution
of the $3s^{2}3p^{6}$ electrons.
In all three figures, the density is shown in the $yz$-plane. The color bar for density is shown on the left.}
\end{figure}

The use of pseudopotentials in a plane-wave calculation defines an effective Hamiltonian,
for which the electronic density is computed. Compared to the all-electron Hamiltonian, 
the contribution of the core electrons are absent in the many-body results.
To illustrate the scale of densities from these
electrons, Fig.~\ref{fig:Cu-psp-eff} compares the computed Cu density from PW-AFQMC 
using two different pseudopotentials:
(i) a Ne-core pseudopotential, the one we use in the full calculation described in the next section, which
leaves 19 valence electrons, including $3s^{2}3p^{6}3d^{10}$  and $4s^{1}$ electrons;
(ii) an Ar-core pseudopotential with 11 valence electrons ($3d^{10}4s^{1}$).
The difference between the two charge density shows the effect of freezing
or retaining the eight $3s^{2}3p^{6}$ electrons; their contribution
to the total density
is  localized around the nuclei, as shown in Fig.~\ref{fig:Cu-psp-eff}(c).

Any direct comparison with the AFQMC densities should be done with the 
same pseudopotential, as is done in Sec.~\ref{sec:benchmark} where we benchmark several DFT functionals.
Assuming good transferability of the pseudopotential, 
we could recover the contribution of the core electrons to the electronic density, at the level of the
independent-electron calculation from which the pseudopotential was generated.  
The core electrons
are frozen in the Kohn-Sham orbitals. The following  provides an estimate of their contributions:
\begin{equation}
\rho_{\mathrm{AE}}^{\mathrm{QMC}}\approx\rho_{\mathrm{psp}}^{\mathrm{QMC}}+(\rho_{\mathrm{AE}}^{\mathrm{DFT}}-\rho_{\mathrm{psp}}^{\mathrm{DFT}})\,,
\end{equation}
where $\rho_{\mathrm{psp}}^{\mathrm{QMC}}$ is our result, and the correction term on the right involves separate DFT calculations,
all-electron (AE) and using the same pseudopotential as in PW-AFQMC (psp).

We use a multiple-projector optimized norm-conserving pseudopotential (ONCVPSP) of Hamann \citep{Hamann_PRB_2013},
whose details are given in Appendix~\ref{sec:psp-supl}.
This pseudopotential was found to allow the use of a lower kinetic
energy cutoff while maintaining excellent accuracy in AFQMC
\citep{Ma_PRB_2017}.

\subsection{\label{subsec:systems}Systems and parameters}

\COMMENTED{
In this work we study Si,
NaCl, and Cu. 
Quantum Espresso \citep{Giannozzi_QE_2009,Giannozzi_QE_2017,Giannozzi_QE_2020}
is used for all DFT simulations other than the all-electron calculations
mentioned in the last section, including generating the trial wave
functions for PW-AFQMC. A multiple-projector optimized norm-conserving
Ne-core pseudopotential (ONCVPSP by D. R. Hamann) \citep{Hamann_PRB_2013}
is used in all three systems by both Quantum Espresso and PW-AFQMC;
this pseudopotential is found to allow the use of a lower kinetic
energy cutoff while maintaining the accuracy, therefore reducing the
computational cost \citep{Ma_PRB_2017}.
}

\noindent \textbf{The covalent-bond crystal silicon (Si).} This Si solid has
a diamond-like structure, with all carbon atoms replaced by silicon
atoms. The primitive cell is a face-centered cubic (FCC) cell, with
an experimental \citep{CODATA_2014} lattice constant of 7.257\,Bohr = 3.840\,\AA.
This FCC cell consists of two Si atoms, located at
$(\pm\frac{1}{8},\pm\frac{1}{8},\pm\frac{1}{8})$.
A commonly used non-primitive cell is the 
cubic cell, with $4\times$  the volume and a lattice constant 
of 5.431\,\AA. 
Both cells, along with the 
$2\times2\times2$
and $3\times3\times3$ multiples of the FCC primitive cells, are used in our
calculations to ensure a finite-size convergence. Densities from FCC
$3\times3\times3$ supercell (54 atoms) are presented, with Baldereschi
mean-value point \citep{Baldereschi_PRB_1973} adopted. We use a Ne-core
pseudopotential which has four valence electrons per silicon atom. The
AFQMC plane-wave kinetic energy cutoff is 25\,Ry.

\noindent \textbf{The ionic crystal sodium chloride (NaCl).} We use
the naturally existing cubic form of NaCl crystal, where the lattice
constant is 5.692\,\AA, taken from the Material's Project \citep{Jain_MatProj_2013,NaCl_MatProj_2021}. 
This cell consists of 4 Na atoms,
at each vertex and face center, and 4 Cl atoms, at the center of each
edge and the bulk center. Although a cubic cell is commonly used,
the actual primitive cell is the $1/4\times$ volume FCC cell, which
comprises of only one Na atom (at lattice points) and one Cl atom
(at bulk center). We use both these cells as well as the $2\times2\times2$
multiple of the primitive FCC cell for finite-size convergence, and
FCC $2\times2\times2$ (eight Na atoms and eight Cl atoms) is used for the
plots below. The Baldereschi mean-value point in the FCC lattice is used.
The kinetic energy cutoff 
is 40\,Ry.

\COMMENTED{ 
In our pseudopotentials here, we use a Ne-core for Cl but He-core for Na, since
its 2s2p semi-core electrons have large overlaps with the 3s electron, and 
neglecting them from the valence was seen to 
cause errors in the equation of state in AFQMC calculations  \citep{Ma_PRB_2017}.
}

\COMMENTED{
were seen to 
(i.e. helium-core) a
Special care need to be taken on the pseudopotential of this
system, as Na's 2s2p electrons are usually considered ``semicore''
and overlaps largely with the one 3s electron, thus neglecting these
``semicore'' electrons would affect calculated properties like equation
of state \citep{Ma_PRB_2017}. Our choice is to keep the eight 2s2p
``semi-core'' electrons as valence electrons instead of freezing
them in the core. The kinetic energy cutoff for both Na and Cl is
40 Ry.
}

\noindent \textbf{The transition metal copper (Cu).} The primitive cell of Cu
is an FCC with all lattice points occupied by copper atoms,
with only one Cu atom per cell. 
We use a 4-atom cubic supercell for most of
the calculations. This cubic cell has an experimental \citep{Ma_PRB_2017,Schimka_JCP_2011} 
lattice constant of 3.59 \AA.  
To characterize the finite-size effect, we also study several larger cells:
FCC $2\times2\times2$ (8 atoms),
BCC $2\times2\times2$ (16 atoms),
and cubic $2\times2\times2$ (32 atoms). 

Cu as a metal requires $k$-point averaging. We use a quasi-random Sobol sequence \citep{Sobol_1967}, with 12
$k$-points in  the cubic $1\times1\times1$
cell. Our calculations are performed with a Ne-core pseudopotential (see Appendix~\ref{sec:psp-supl}) and
a kinetic energy cutoff of 64\,Ry.

\section{\label{sec:density}Electronic density from AFQMC}

In this section we present the final charge densities computed with PW-AFQMC for Si, NaCl, and Cu.
For each system, the results are presented on a selected 
high-symmetry plane, and then along a path of one-dimensional (1D) line cuts. 
In all three solids we plot the final densities from a cubic $1\times1\times1$
supercell 
regardless of the actual supercell size used in calculation
(see Sec.~\ref{subsec:systems}).
Numerical values are provided as supplemental information online, see Ref. \citep{Chen_GitHub_2020}.

\begin{figure}
\includegraphics[width=1\linewidth]{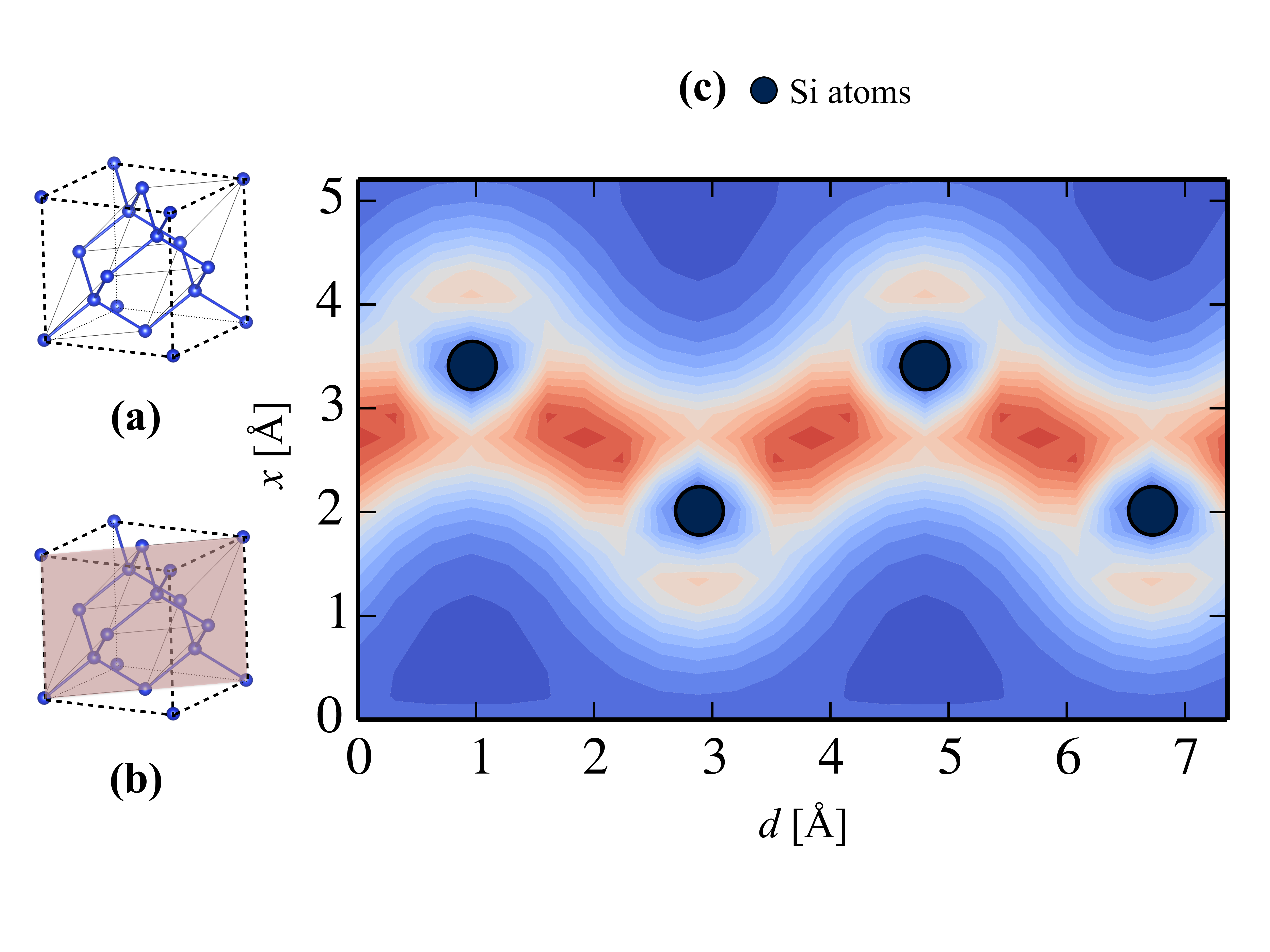}\caption{\label{fig:Si-plane-plot}
Charge density of Si from PW-AFQMC. The lattice structure shown
in (a), and (b) illustrates the $(01\bar{1})$ plane  ($y=z$) for which the density is plotted. In (c),
higher density is given in red and lower density in blue.
The vertical axis is the Cartesian $x$-value, while the horizontal axis gives $d=\frac{1}{\sqrt{2}}(y+z)$.
}
\end{figure}

\begin{figure}
\includegraphics[width=1\linewidth]{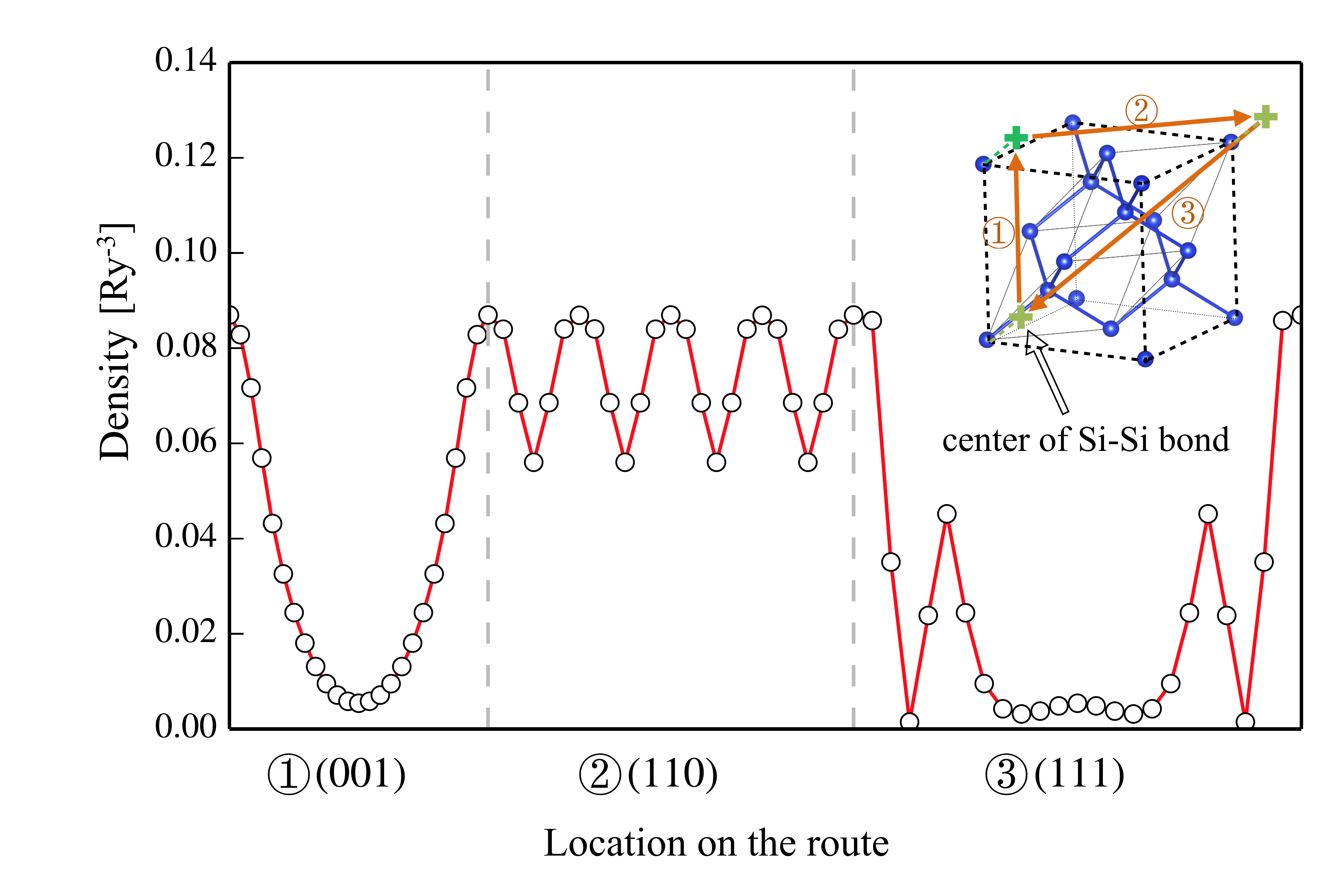}\caption{\label{fig:Si-route-plot}
Charge density of Si from PW-AFQMC, plotted on a high-symmetry path. 
The starting and ending point of the route is the origin of the simulation cell, which is
located at the center of two nearest neighbor
Si atoms. The route then goes through three segments in $\langle001\rangle$,
$\langle110\rangle$, and $\langle111\rangle$ directions, as indicated in the inset.} 
\end{figure}

In Si, we plot the density in the $(01\bar{1})$
plane in Fig.~\ref{fig:Si-plane-plot}.
The signature of the Si-Si covalent bond is evident, as 
a concentration of valence electrons is seen on a line connecting the nearest neighbor Si atoms.
Note that there are small bumps
located in the vicinity of each Si atom and complementing the two bonds in the plane, which
are contributions from two Si-Si bonds pointing out of the plane, reflecting the nature of the  $sp^{3}$  hybridization.
We also plot the charge density along line cuts in  Fig.~\ref{fig:Si-route-plot}, following the route
 $O-$$\langle001\rangle$$-O{'}-$$\langle110\rangle$$-O{''}-$$\langle111\rangle$$-O$,
 which forms a triangle as illustrated in the inset.
 The origin  $O$ is  taken to be the high-symmetry middle point
between two neighboring Si atoms, while $O{'}$ and  $O{''}$ are translated from  $O$ by lattice constants along the direction connecting them.

\begin{figure}
\includegraphics[width=1\linewidth]{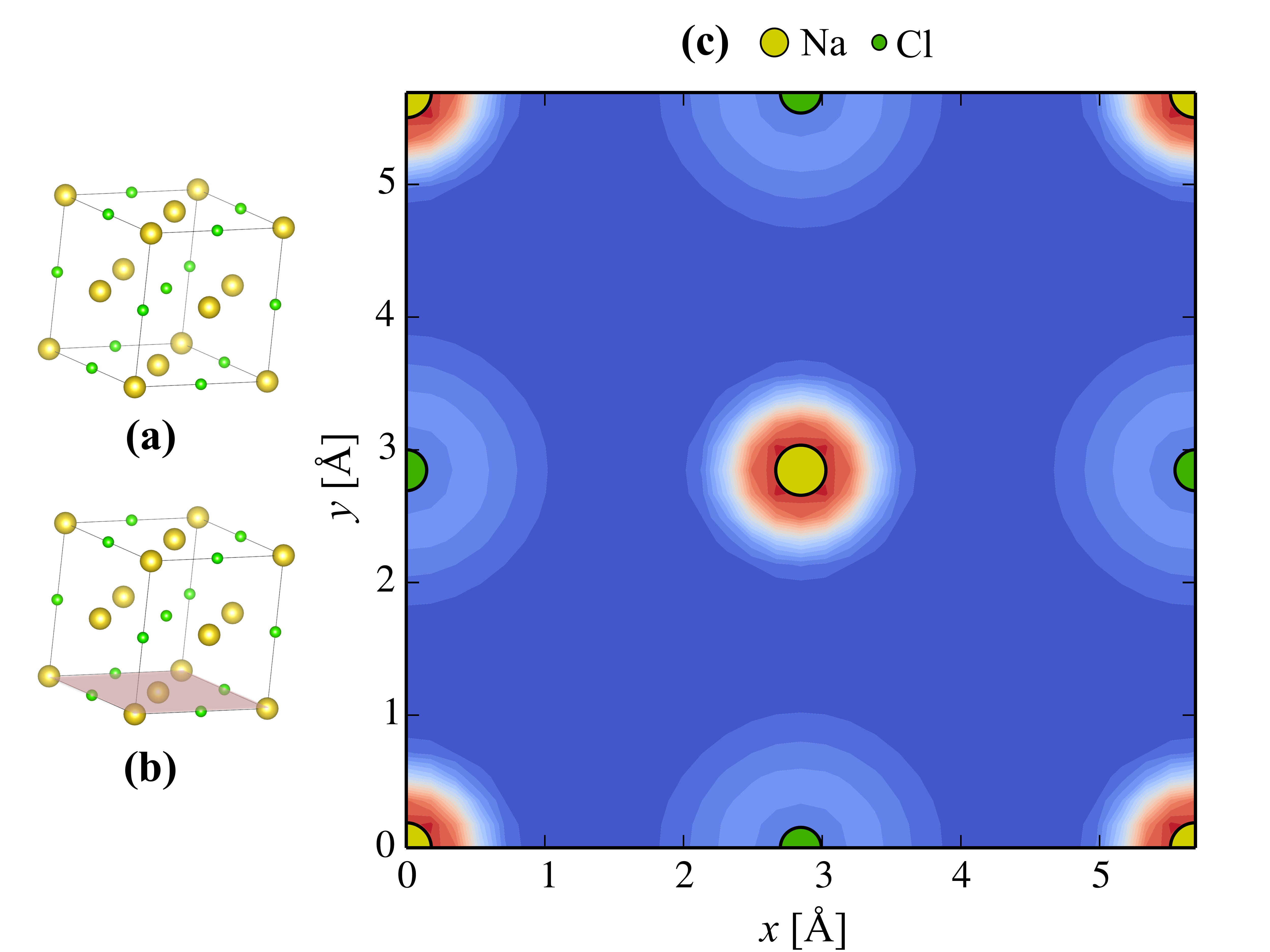}\caption{\label{fig:NaCl-plane-plot}
Charge density of NaCl from PW-AFQMC. 
The layout is the same as Fig.~\ref{fig:Si-plane-plot}. 
The density is plotted in the $xy$-plane.
}
\end{figure}

\begin{figure}
\includegraphics[width=1\linewidth]{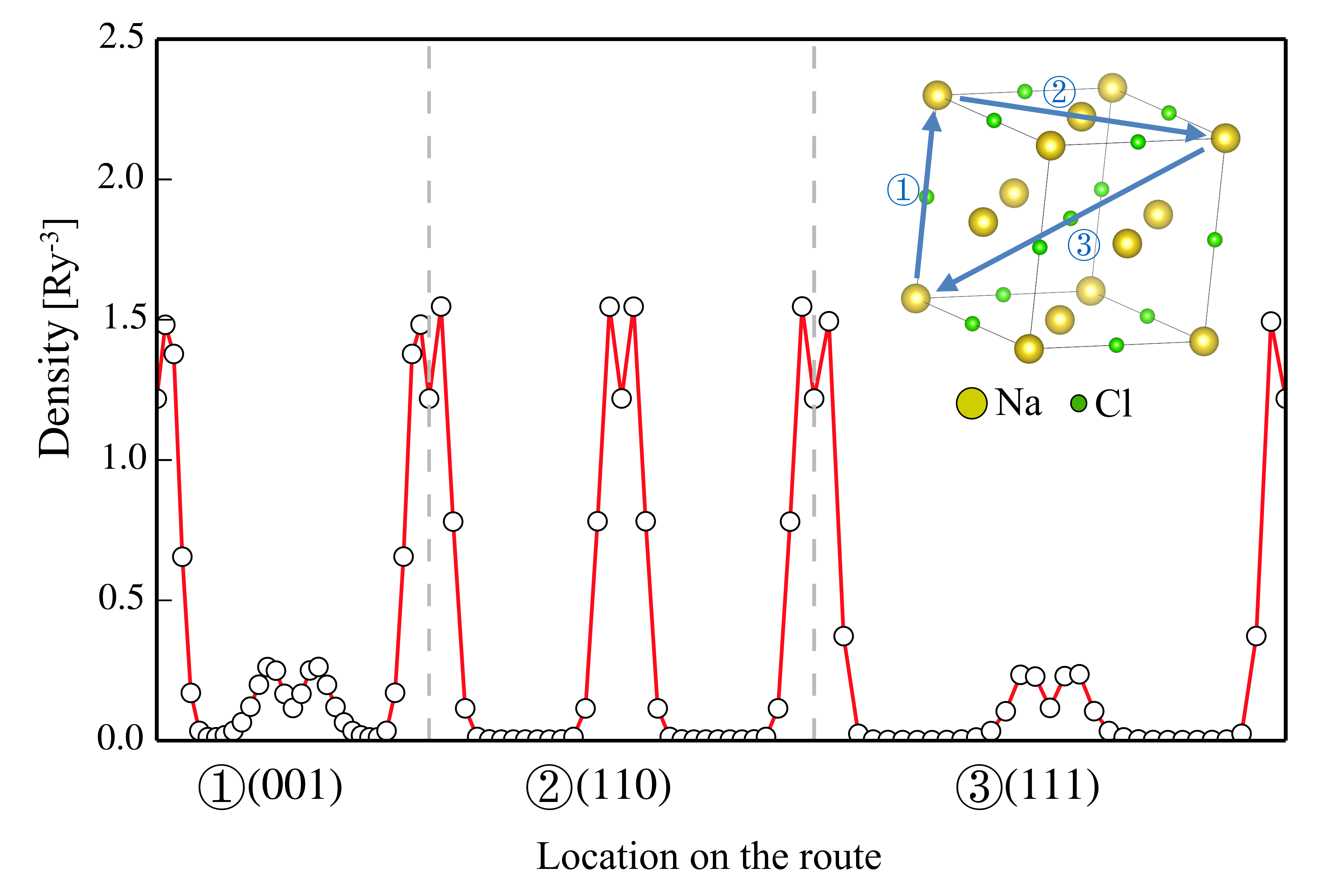}\caption{\label{fig:NaCl-route-plot}
Charge density of NaCl from PW-AFQMC, plotted on
a high-symmetry path. 
The layout is the same as Fig.~\ref{fig:Si-route-plot}.  
\COMMENTED{
Plot of NaCl PW-AFQMC density on a high-symmetric
1D route. (Route illustrated as a subplot: the starting and ending
point of the route is the origin of the calculation cell, located
at a Na atom. The route then goes through three segments on $\langle001\rangle$,
$\langle110\rangle$, and $\langle111\rangle$ directions.)}
}
\end{figure}

\begin{figure}
\includegraphics[width=1\linewidth]{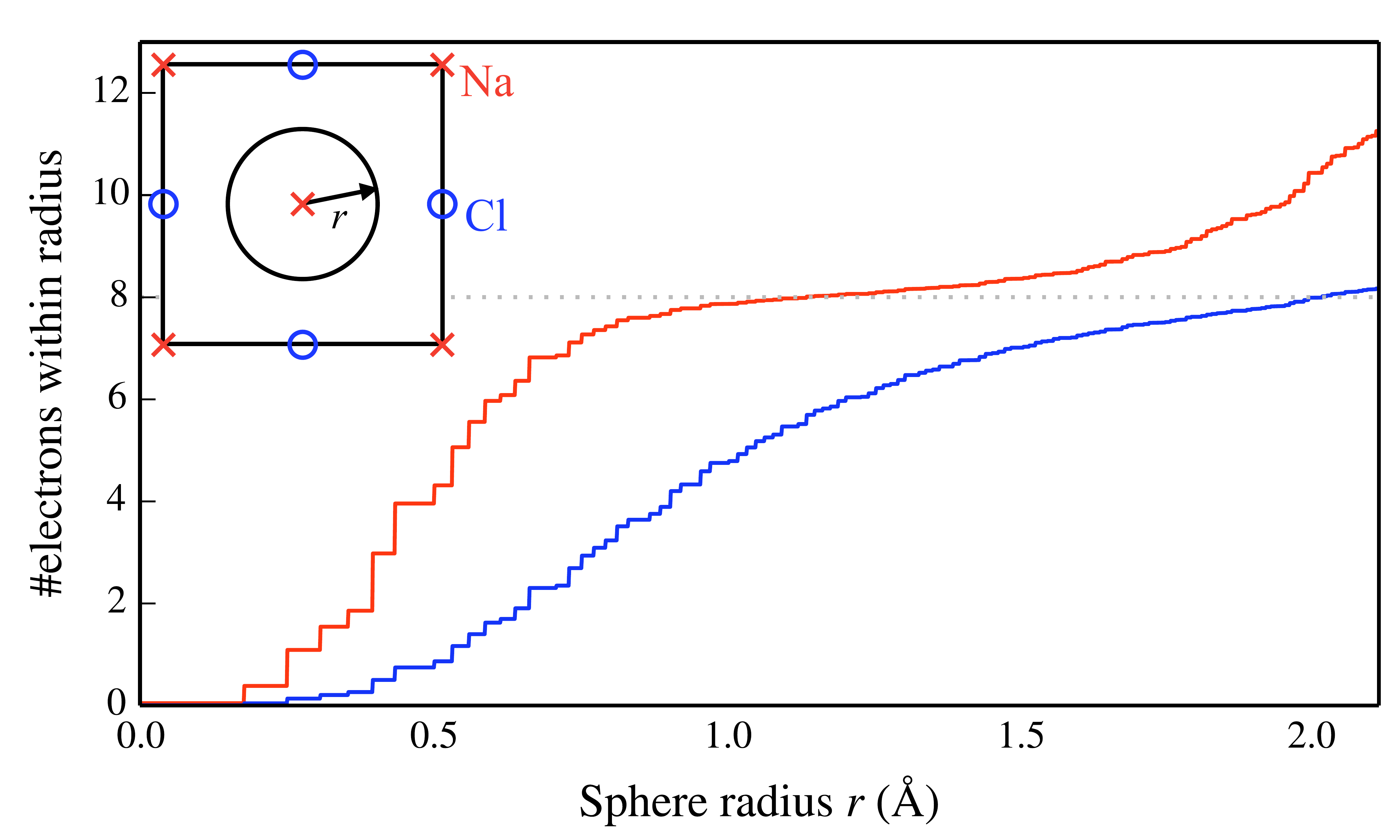}\caption{\label{fig:NaCl-rad-elec}
Number of electrons within a sphere of radius $r$ centered at either a Na or a Cl  nucleus 
in the computed PW-AFQMC density.
Red curve is for Na and blue curve is for Cl.}

\end{figure}

We next present the results for NaCl in a similar fashion. 
In Fig.~\ref{fig:NaCl-plane-plot}, the density is shown on  the 
$xy$-plane ($z=0$). 
Then in Fig.~\ref{fig:NaCl-route-plot}, we plot the density along the triangular path, 
following the route $O-$$\langle001\rangle$ $-O{'}-$$\langle110\rangle$$-O{''}-$$\langle111\rangle$$-O$,
with the origin $O$ taken to be a Na atom.
Figure \ref{fig:NaCl-rad-elec} shows the density from a different perspective, illustrating the ionic nature of NaCl.
Within a sphere of radius 1\,\AA \ centered at a Na atom, 
the integrated charge density is $\sim 8$ electron, 
consistent with a Na$^+$ (with our He-core pseudopotential). 
This integrated charge density remains saturated  with increasing radius, 
until the sphere reaches the vicinity of the nearest Cl atom.
Around a Cl atom, this integrated density also approaches $\sim 8$ electron 
when the radius of the sphere is around 2\,\AA, consistent with a Cl$^-$ ion (Ne-core  pseudopotential).

\begin{figure}
\includegraphics[width=1\linewidth]{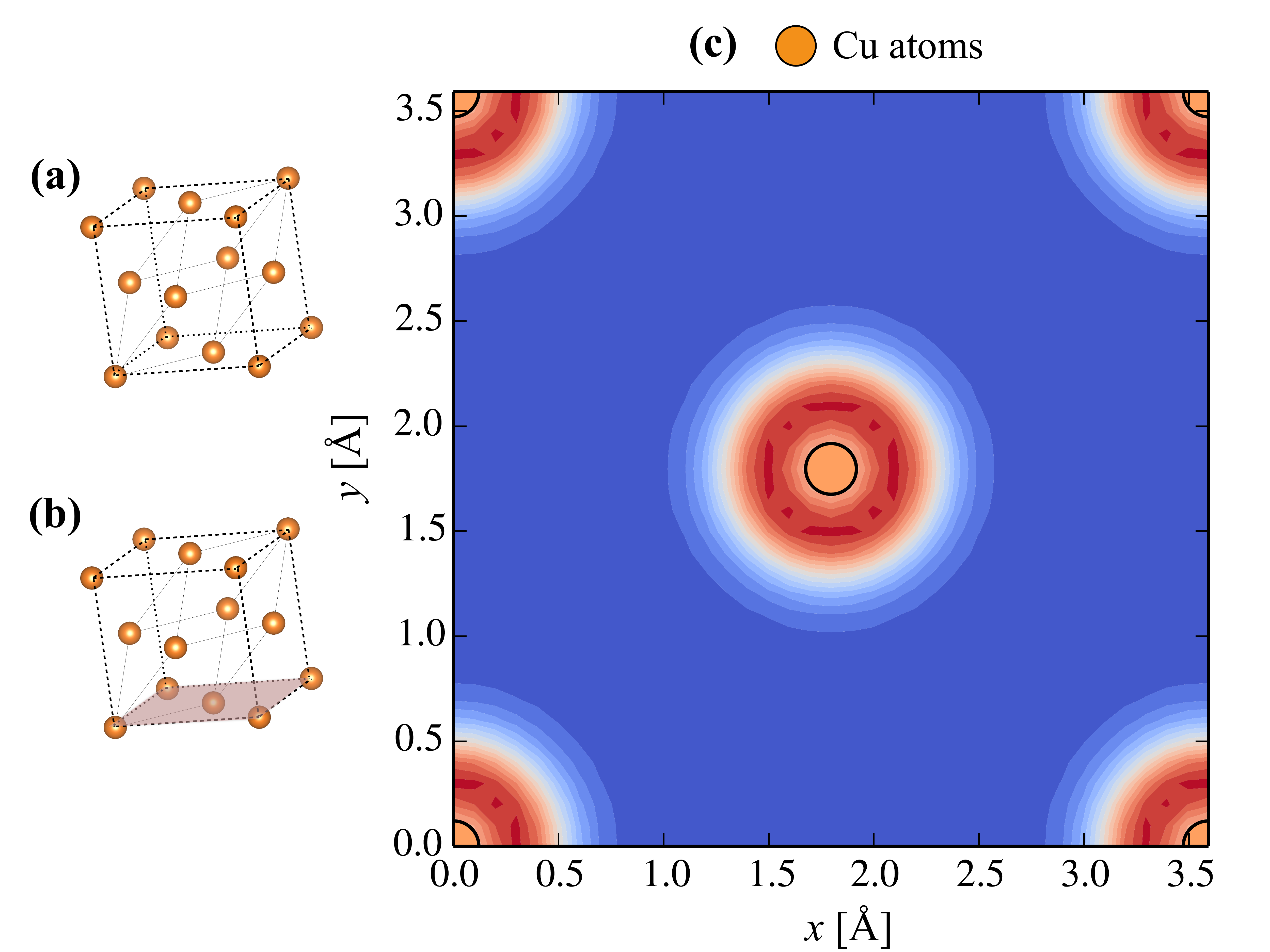}\caption{\label{fig:Cu-plane-plot}
Charge density of Cu from PW-AFQMC. 
The layout is the same as Fig.~\ref{fig:Si-plane-plot}. 
The density is plotted in the $xy$-plane.
}
\end{figure}
\begin{figure}
\includegraphics[width=1\linewidth]{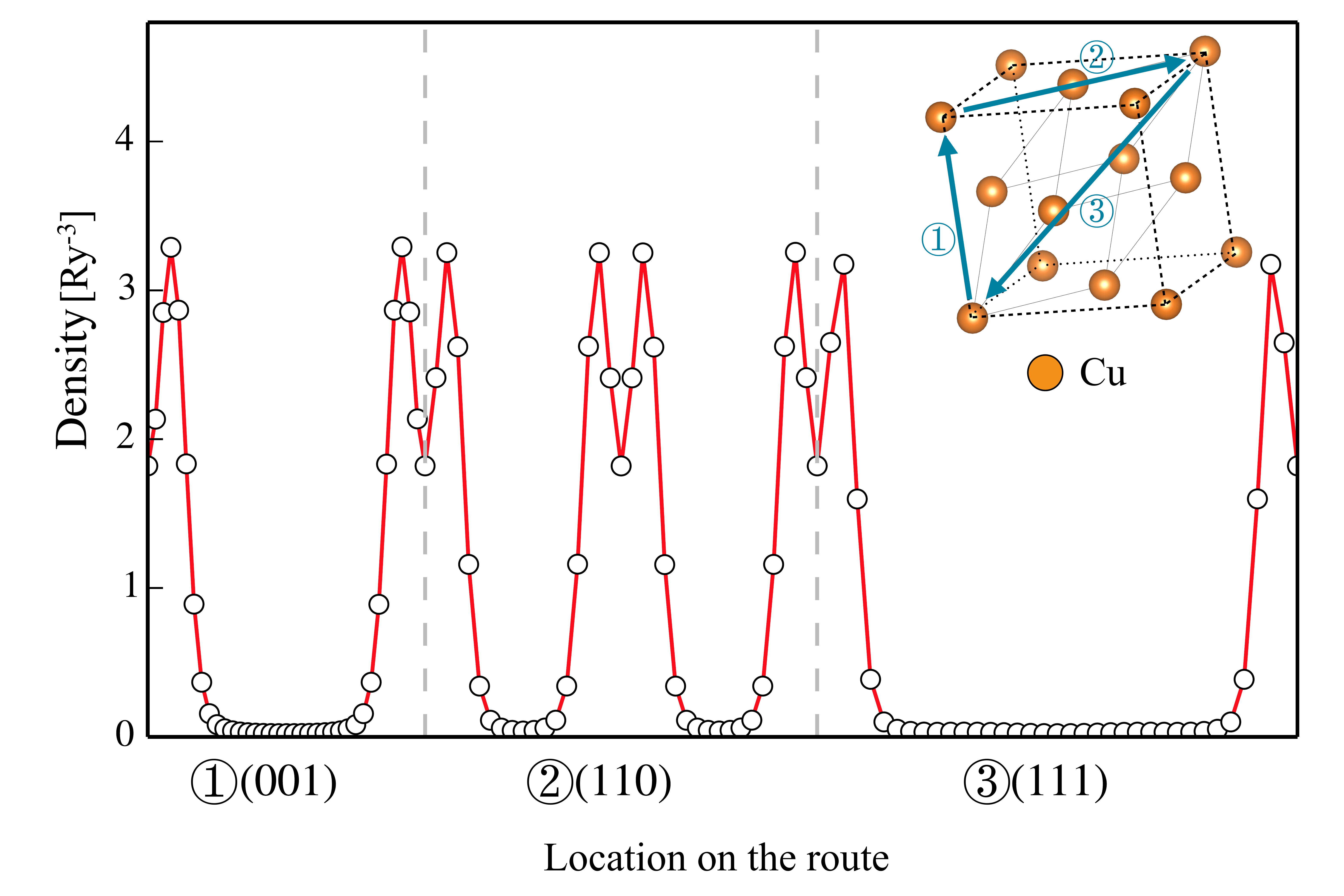}\caption{\label{fig:Cu-route-plot}
Charge density of Cu from PW-AFQMC, plotted on
a high-symmetry path. 
The layout is the same as Fig.~\ref{fig:Si-route-plot}.  
}
\end{figure}

\begin{figure}
\includegraphics[width=1\linewidth]{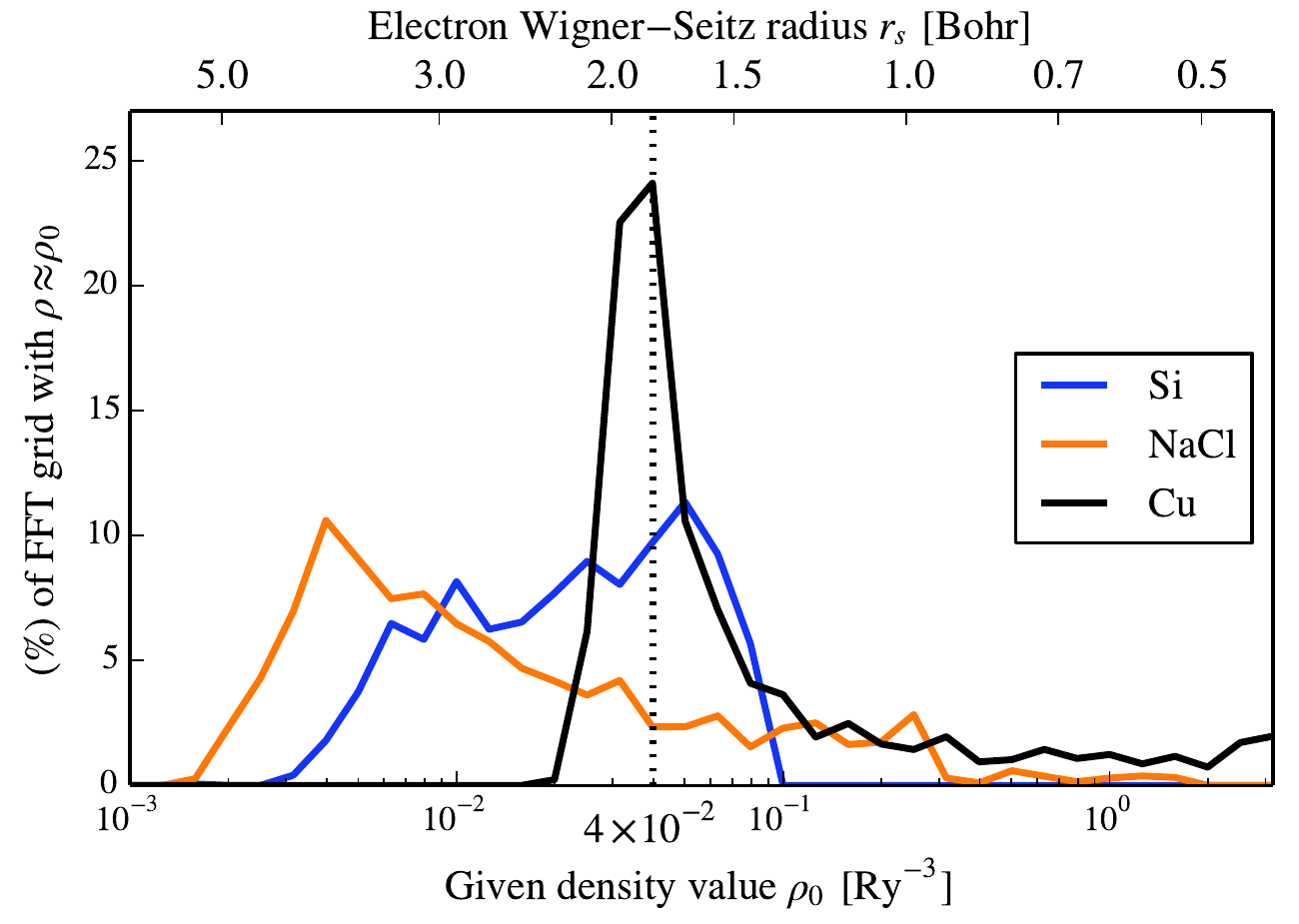}\caption{\label{fig:densval-stat}Distribution of the electron density in 
different solids. Shown are the 
percentage of the real-space FFT
grid points with density around each given value $\rho_{0}$, for Si,
NaCl, and Cu from the PW-AFQMC densities.
Corresponding values of the electron Wigner-Seitz radius $r_s$ are marked on top.
Note the logarithmic 
scale of the horizontal axis.
}
\end{figure}

Results for Cu are shown in Figs.~\ref{fig:Cu-plane-plot} and  \ref{fig:Cu-route-plot}.
The charge density in Cu bears a resemblance to that of NaCl. However the density scales are very different in 
Fig.~\ref{fig:Cu-route-plot}, which obscures the significant density in between Cu atoms. The distinction between 
this metallic system and the ionic crystal NaCl, and the semiconductor Si can be seen more clearly in Fig.~\ref{fig:densval-stat},
in which we plot the distribution of electron densities in each solid. The distribution is characterized by the 
percentage of the real-space FFT grid points having a given
charge density value. 
(It should be noted that this is in presence of the pseudopotential and without accounting for the core electrons.)
We see that a pronounced peak is present in Cu at around  $\rho=0.04[\mathrm{Ry}^{-3}]$,
with 55\% of FFT grid points in Cu share a density value
of $(4\pm1)\times10^{-2}$ $[\mathrm{Ry}^{-3}]$, 
corresponding to an electron Wigner-Seitz radius $r_s$ of $1.7-2.0$ Bohr. 
This is consistent with the notion of a metallic system being characterized as a ``uniform electron gas'',
and is not seen 
in NaCl or Si.
In particular, NaCl shows a significant concentration in a low-density region, while Si features a broader peak.

\section{\label{sec:benchmark}Benchmark DFT functionals}

\subsection{\label{subsec:density-benchmark}Comparison of the electronic densities from several functionals with AFQMC results}

Based on a variety of benchmark studies (see, for example, Refs.~\cite{LeBlanc_PRX_2015,Williams_PRX_2020,Motta_PRX_2017}),
the AFQMC results are expected to be highly accurate. 
The density results we have presented from AFQMC can serve as a 
reference in these systems.
We next carry out a comparative study of the computed densities from several popular DFT functionals, including:
(i) LDA, the local-density approximation 
by Perdew-Zunger (PZ) \citep{Perdew_LDA_1981};
(ii) PBE, a generalized-gradient approximation (GGA) functional 
by Perdew-Burke-Ernzerhof \citep{Perdew_PBE_1996}; 
(iii) PBEsol, a revised GGA functional for solids \citep{Perdew_PBEsol_2008};
(iv) PBE0, a hybrid functional between PBE and Hartree-Fock \citep{Perdew_PBE0_1996};
(v) B3LYP, a widely used hybrid functional in quantum chemistry \citep{Stephens_B3LYP_1994}. 
In each DFT calculation, we use the same pseudopotential as in the PW-AFQMC calculations, 
namely the multiple-projector ONCVPSP generated with LDA. In other words, we adopt the philosophy of 
viewing the pseudopotential as defining an effective Hamiltonian, and ask how each functional treats this effective Hamiltonian as 
compared to AFQMC.


\begin{figure}
\includegraphics[width=1\linewidth]{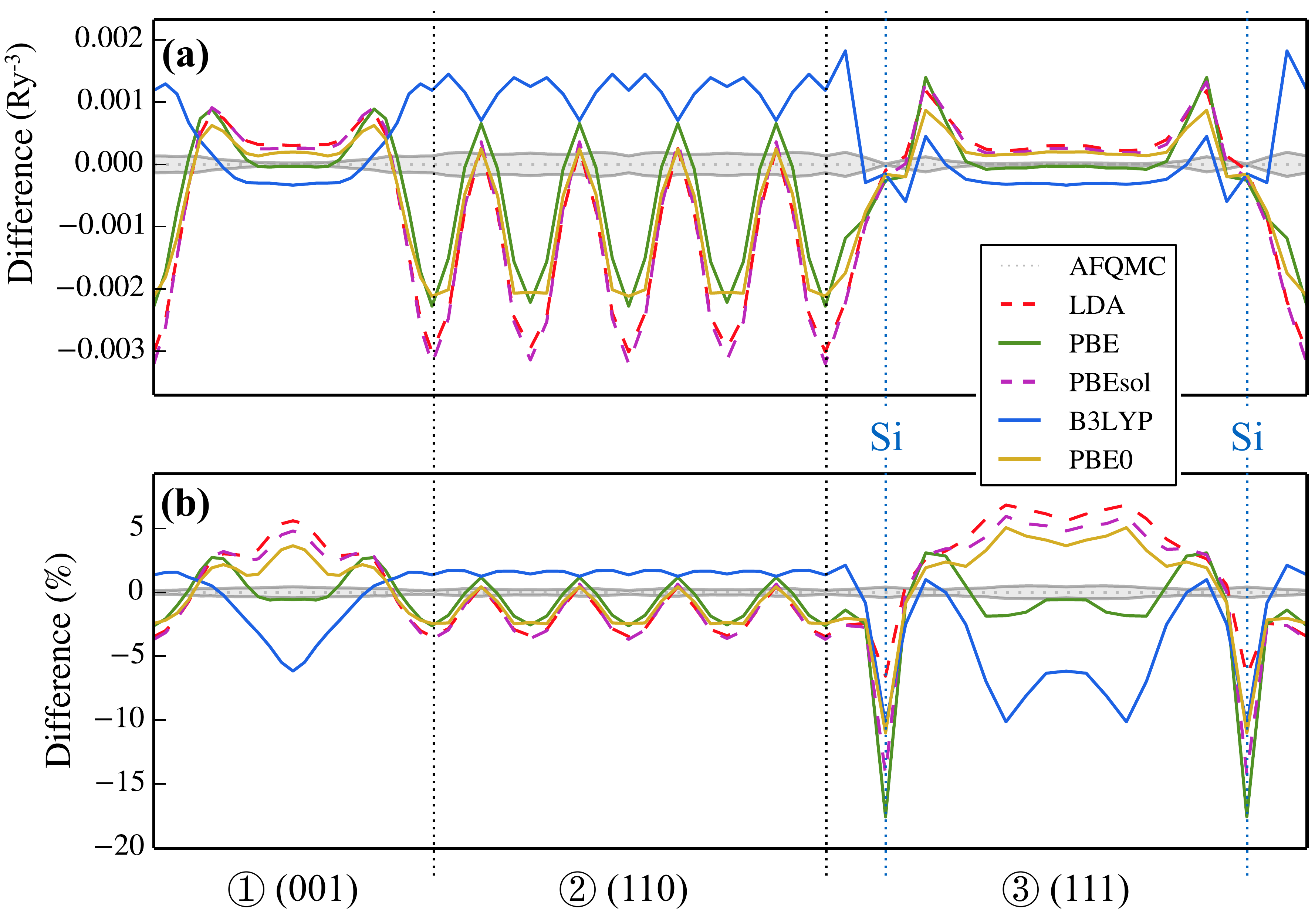}
\caption{\label{fig:Si-DFT-benchmark}Benchmark of five DFT exchange-correlation functionals
against the PW-AFQMC density
in Si. (a) The difference between the DFT and AFQMC densities and (b)
 the relative errors as a percentage are plotted along 
 the same line cut as in Sec.~\ref{sec:density}.
The gray shade indicates the AFQMC statistical error bar.
Blue vertical lines mark the position of the Si atoms.}
\end{figure}

\begin{figure}
\includegraphics[width=1\linewidth]{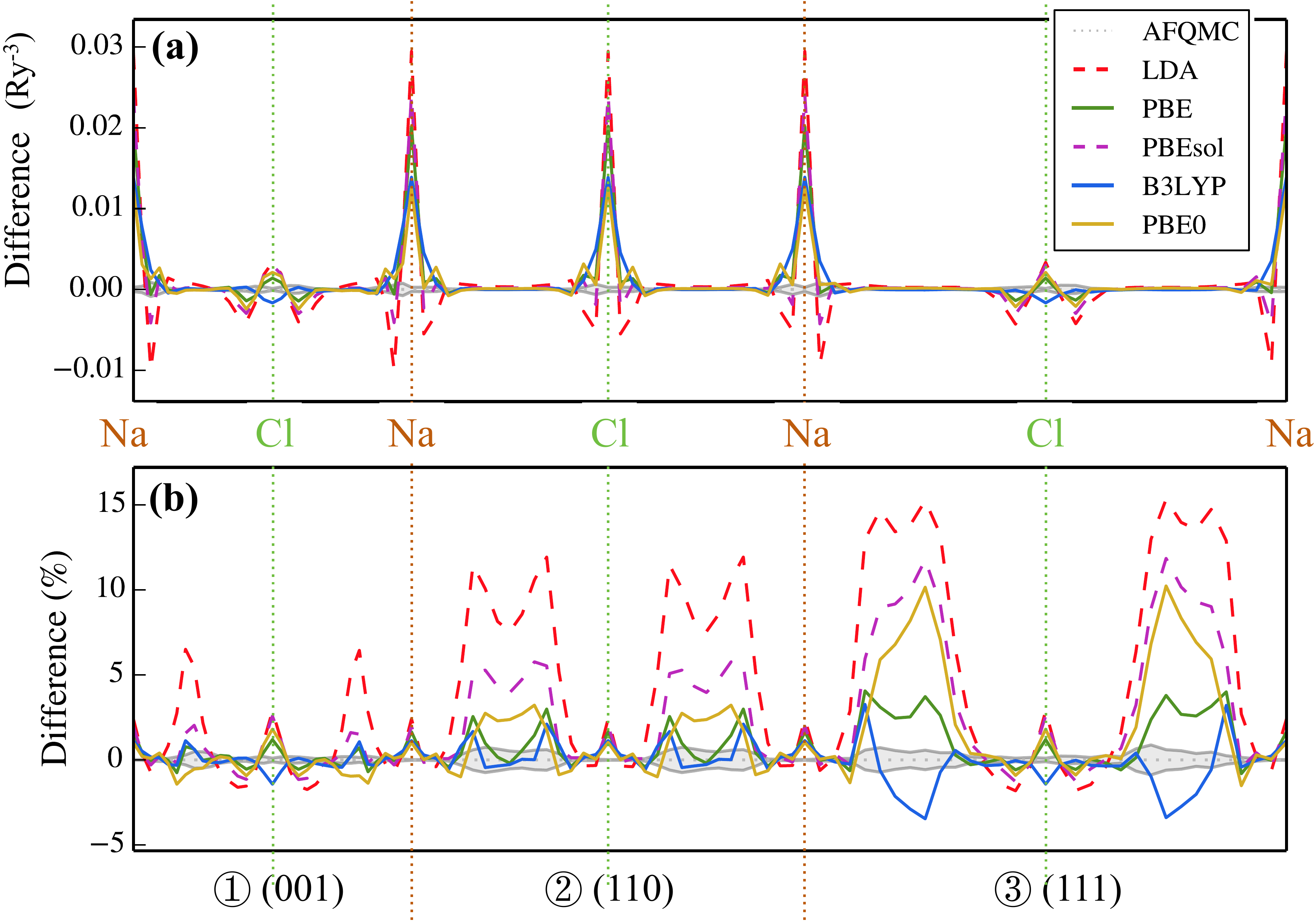}
\caption{\label{fig:NaCl-DFT-benchmark}Same as Fig. ~\ref{fig:Si-DFT-benchmark},
but for NaCl. Brown and green vertical lines mark the positions of
Na and Cl atoms, respectively.
}
\end{figure}

\begin{figure}
\includegraphics[width=1\linewidth]{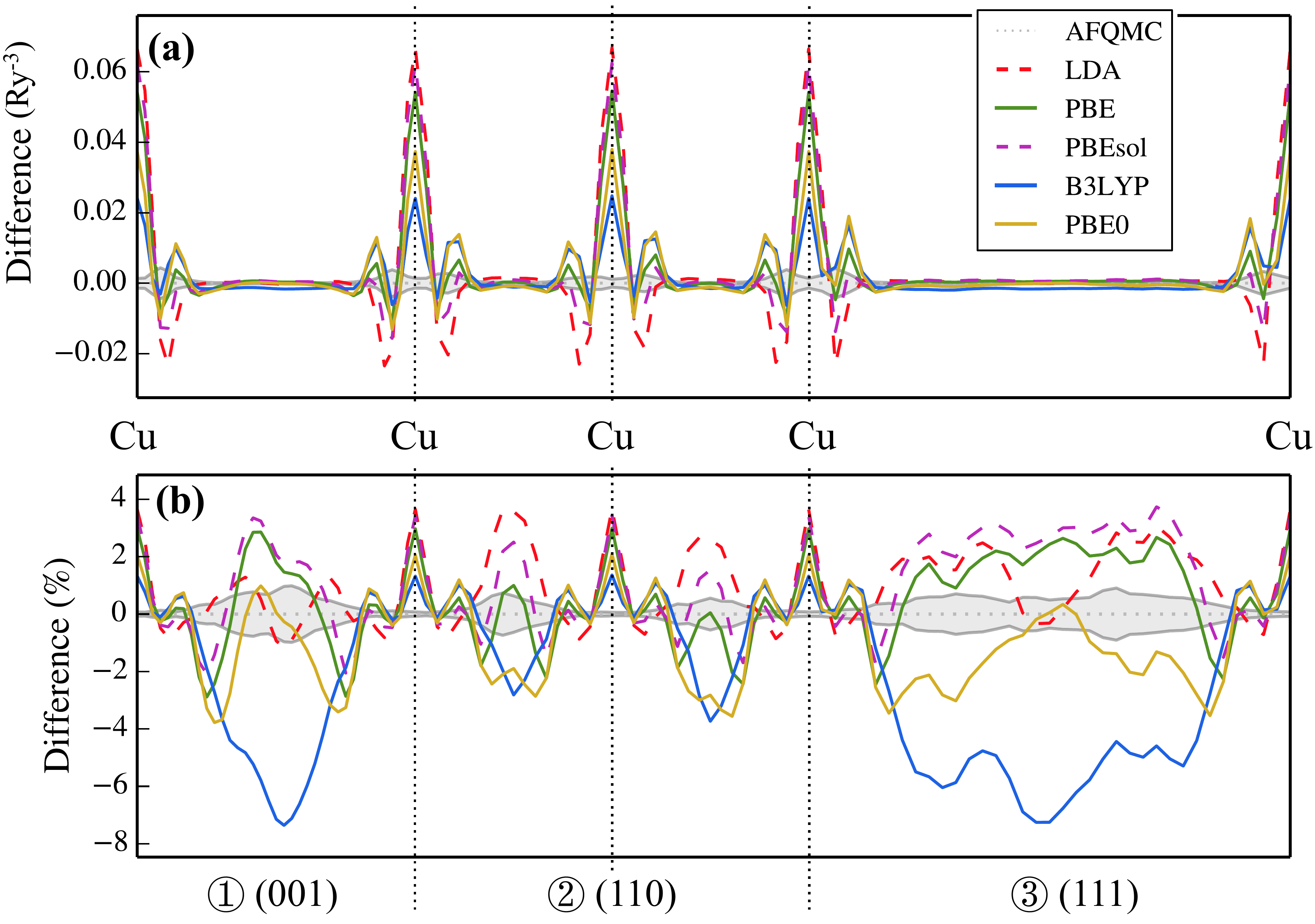}

\caption{\label{fig:Cu-DFT-benchmark}Same as Fig. ~\ref{fig:Si-DFT-benchmark},
but for fcc-Cu. Black vertical lines mark the positions of the Cu
atoms.}
\end{figure}

In Figs. \ref{fig:Si-DFT-benchmark},  \ref{fig:NaCl-DFT-benchmark}, and  \ref{fig:Cu-DFT-benchmark}, we present the result for Si, NaCl, and Cu, respectively,
by plotting the densities from the five different DFT functionals
against the reference AFQMC density. Density differences and percentage discrepancies
are shown in two different panels, following the same line cuts as used in Figs.~\ref{fig:Si-route-plot}, \ref{fig:NaCl-route-plot}, and \ref{fig:Cu-route-plot},
respectively. The AFQMC statistical error bars are given by the shades. 
Additional details on the benchmark are provided in Appendix \ref{sec:benchmark-supl}.
 
For Si, PBE and PBE0 perform very well, with average percentage errors of around 2\%.
B3LYP shows a comparable performance 
along the route, but is less accurate 
for the entire cell.
LDA and PBEsol show larger errors in general. We see correlations in the location and amount of the errors between all functionals, except for B3LYP whose behavior is more different from the other functionals. 
In NaCl, the most significant errors occur between Na and Cl atoms, along the diagonal direction.
Except for B3LYP, all functionals tend to give a density that is consistently too high in these regions, well outside the AFQMC statistical error.
The errors from the five functionals 
all show some correlation, with B3LYP giving the smallest errors in these regions, 
followed by PBE and PBE0. 
In Cu, the behavior of the errors is more subtle. The absolute and relative errors in the plot tell somewhat different stories.
B3LYP yields the smallest integrated absolute error along the chosen path, but gives the largest percentage error, and also the second 
largest integrated absolute error over the entire cell. 


\subsection{\label{subsec:discussions}Accuracy of a functional: density vs.~total energy}

\begin{table}[b]
\caption{\label{tab:EOS-XC}
Equilibrium lattice constant ($a_{\mathrm{eq}}$)
and bulk modulus ($B_{0}$) computed from several DFT functionals, compared to experimental results in Si and NaCl.
The last two rows give the results from one-shot calculations of the EOS using PBE exchange-correlation functional but densities
(wave functions) obtained from B3LYP and LDA, respectively.
}
\begin{ruledtabular}
\begin{tabular}{c|c|cc|cc}
\multirow{2}{*}{XC} & \multirow{2}{*}{\shortstack{Input\\Density}} & \multicolumn{2}{c|}{NaCl} & \multicolumn{2}{c}{Silicon}\tabularnewline
\cline{3-6} 
 &  & $a_{\mathrm{eq}}$ (\AA) & $B_{0}$ (GPa) & $a_{\mathrm{eq}}$ (\AA) & $B_{0}$ (GPa)\tabularnewline
\hline 
Exp. & & 5.640\citep{Barrett_JACS_1954} & 24.42\citep{Combes_JOSA_1951} & 5.431\citep{Kittel_SSPhys_1996} & 98.8\citep{Kittel_SSPhys_1996}\tabularnewline
\hline 
LDA & \multirow{4}{*}{(same)} & 5.46 & 32.21 & 5.395 & 95.19\tabularnewline
B3LYP &  & 5.58 & 27.98 & 5.382 & 96.40\tabularnewline
PBEsol &  & 5.60 & 26.06 & 5.394 & 94.89\tabularnewline
PBE &  & 5.672 & 24.28 & 5.406 & 92.04\tabularnewline
\hline 
\multirow{2}{*}{PBE} & B3LYP & 5.681 & 23.27 & 5.408 & 90.65\tabularnewline
 & LDA & 5.682 & 22.93 & 5.404 & 91.53\tabularnewline
\end{tabular}
\end{ruledtabular}
\end{table}

Although charge density plays a vital role in DFT,
the accuracy of the density from a particular exchange-correlation functional does not seem to provide an unambiguous 
measure of 
the quality of 
the functional.
We illustrate this point below following the benchmark results on Si and
NaCl in the previous subsection.
  
We perform equation of state (EOS) calculations in these solids to obtain the 
equilibrium lattice constant and bulk modulus by fitting the total energies using the Murnaghan equation \citep{Murnaghan_PNAS_1944}.
The results are tabulated in  Table \ref{tab:EOS-XC}, together with the experimental values. 
In NaCl, PBE yields the best results for both the equilibrium lattice constant and the bulk modulus. 
However, we recall that B3LYP is the one that yielded the best density.
In Si, B3LYP shows the largest error in the lattice constant and the best accuracy in bulk modulus, 
while PBE is at the opposite, with the smallest error in lattice constant and the largest in bulk modulus, although the margins are all rather small here. 
On the other hand, the result from the overall electronic density shows no ambiguity and indicates that PBE performs the best, as discussed in Sec.~\ref{subsec:density-benchmark}.
\COMMENTED{
They show that, although B3LYP performs well, PBE and PBEsol yield results that are having better agreement with the experiment in NaCl;
for Si, B3LYP and PBE/PBEsol alternates in quality for the lattice constant and the bulk constant.
These findings seem to be inconsistent with the results of the density benchmark.
 }
 
To reconcile these inconsistent behaviors in density versus total energy, 
we examine the effect of the exchange-correlation energy separately from the density. 
We use the PBE functional to perform a one-shot total energy calculation by feeding it  a wave function from B3LYP (or LDA). 
The wave function is obtained from a converged self-consistent 
B3LYP calculation, by taking the occupied orbitals. Thus the wavefunction 
produces the electronic density from the fully self-consistent B3LYP calculation. 
The resulting EOS using total energies computed from this procedure is re-analyzed, and shown in the 
bottom two rows, for B3LYP and LDA wave functions respectively. 
We see that the results in general 
exhibit only minor variations from the PBE results.
In the case of LDA, the improvement is dramatic
in NaCl, and mixed in Si, with the bulk modulus becoming worse as it moves towards the PBE result (which also occurs with B3LYP density). This shows that the accuracy of the electronic 
density is secondary in the performance of an XC functional for computing the EOS,
and the
primary factor affecting overall performance 
in these solids  is the exchange-correlation energy.

\section{\label{sec:conclusion}Conclusion}

We have introduced into the plane-wave AFQMC framework  the calculation of observables using 
a path-restoration back-propagation technique,
which is applied to compute the charge density in three typical solids
with different crystal bonding mechanisms,  the covalent-bond crystal
Si, the ionic-bond NaCl, and the transition metal fcc-Cu.
The results provide 
highly accurate 
\textit{ab initio} many-body electronic densities in these systems.
We compared these results against several of the most popular density functionals.
In general, the densities produced by these functionals agree quite well with the PW-AFQMC results. The discrepancies from the different functionals
are quantified, and our results can be used for future benchmarks of other computational methods.
Additionally, the PW-AFQMC electronic densities may help with the development of improved density functionals.

Besides charge density, the back-propagation technique extends  easily to other quantities, including
interatomic forces, which can be used to perform accurate molecular dynamics
and geometry optimization,
as well as to calculate phonon properties
and access thermodynamic properties of solids. 
Work is on-going along these lines.

\begin{acknowledgments}
We thank D. Ceperley, M. Holzmann, H. Krakauer, E. J. Walter and J. Lee for useful discussions.
We are grateful to L. Reining for discussions and for pointing out an error in an early version of the manuscript.
S.C. would like to thank the Center for Computational Quantum Physics, Flatiron Institute 
for support and hospitality. 
We also acknowledge support from the U.S. Department of Energy (DOE) 
under Grant No. DE-SC0001303.
F.M. is supported by the National Natural Science Foundation of China
under Grant No. 11674027.
The authors thank William \& Mary Research Computing and 
Flatiron Institute Scientific Computing Center for computational resources and technical support. 
The Flatiron Institute is a division of the Simons Foundation.
\end{acknowledgments}

\appendix

\section{\label{sec:psp-supl}More information on the pseudopotential}

Our pseudopotentials for Si, NaCl, and Cu are constructed with 
Hamann's multiple-projector optimized norm-conserving pseudopotential (\textsc{oncvpsp}) code  \citep{Hamann_PRB_2013}, 
using LDA as the exchange correlation functional. 
The local reference in \textsc{oncvpsp} is a smooth continuation of the all-electron potential from the minimum radius cutoff $r_c$ to 0, using a smooth polynomial ($lloc=4$ in \textsc{oncvpsp} input).
For compatibility with AFQMC, the nonlinear core correction is turned off in pseudopotential generations. 

For Si, the kinetic energy cutoff is 25 Ry and 
the electron configuration is $1s^{2}$$2s^{2}$$2p^{6}$$3s^{2}$$3p^{1.6}$$3d^{0.4}$.
The first three orbitals are taken as core orbitals.
We use three projectors $l=0,1,2$, all with the same core radius $r_c$ of 2.04 Bohr.

For NaCl, the kinetic energy cutoff is 40 Ry. 
In Na, the electron configuration 
is $1s^{2}$$2s^{2}$$2p^{6}$$3s^{1}$, 
with only $1s$ orbital as core, 
due to the fact that
the $2s2p$ semi-core electrons of Na have large overlaps with the $3s$ electron, and 
neglecting them from the valence was seen to 
cause errors in the equation of state in AFQMC calculations  \citep{Ma_PRB_2017}.
Core radii $r_c$ for the three projectors $l=0,1,2$ are 1.82, 2.27, 2.27 Bohr, respectively.
%
In Cl, the electron configuration is $1s^{2}$$2s^{2}$$2p^{6}$$3s^{2}$$3p^{5}$, 
and the first three orbitals are taken as core. 
The core radii for the three projectors ($l=0,1,2$) are 2.72, 2.72, and 3.01 Bohr, respectively.

For Cu, we have used two different pseudopotentials, with Ne-core and Ar-core, both using a kinetic energy cutoff of 64\,Ry.
The Ne-core pseudopotential, which is the primary one used in the paper,
has an electron configuration of
$1s^{2}$$2s^{2}$$2p^{6}$$3s^{2}$$3p^{6}$$3d^{9.1}$$4s^{0.5}$$4p^{1.4}$.
Core radii are 1.60, 1.97, and 1.97 Bohr for the three projectors; the corresponding wave-vector cutoffs are 7.0, 7.75 and 8.0\,Ry.
%
 For the Ar-core pseudopotential, the electron configuration is
$1s^{2}$$2s^{2}$$2p^{6}$$3s^{2}$$3p^{6}$$3d^{9}$$4s^{1}$$4p^{1}$,
with the first five orbitals being the core orbitals.
Core radii for $l=0,1,2$ are 2.30, 2.30, and 2.10 Bohr, respectively. The wave-vector cutoff (8 Ry) 
is the same for all projectors.

\section{\label{sec:benchmark-supl}Supplementary Data on Exchange-Correlation
Functional Benchmarks}

In Tables \ref{tab:Si-dens-benchmark}, \ref{tab:NaCl-dens-benchmark},
and \ref{tab:Cu-dens-benchmark}, we list the 
mean absolute error (MAE) in the density
of multiple DFT exchange correlation functionals with respect to the
near-exact AFQMC density, on the high-symmetry triangular route $O-$$\langle001\rangle$
$-O_{001}-$$\langle110\rangle$$-O_{111}-$$\langle111\rangle$$-O$
and the full three-dimensional (3D) real-space grid, for Si, NaCl, and Cu, respectively.
The formula for this MAE is 
\[
\bar{|\Delta\rho|}=\frac{1}{N_{G}}\sum_{g\in\{G\}}|\Delta\rho_{g}|\,,
\]

\noindent where $\{G\}$ is the set of real-space grid points, located
on the triangular route or the full 3D real-space grid; $N_{G}$ is
the number of grid points in $\{G\}$; and $\Delta\rho_{g}=\rho_{\mathrm{XC}}-\rho_{\mathrm{AFQMC}}$
for a given exchange-correlation functional ``XC.'' 
For mean percentage 
difference a similar formula is used,
\[
\Delta\rho_{g}=100\,(\%)\times\frac{\rho_{\mathrm{XC}}-\rho_{\mathrm{AFQMC}}}{\rho_{\mathrm{AFQMC}}}\,.
\]

\begin{table*}[b]
\caption{\label{tab:Si-dens-benchmark} MAE and  mean percentage difference 
of DFT densities in Silicon with LDA, PBE, PBEsol, B3LYP, and PBE0 exchange-correlation functionals,
compared to AFQMC results (in bold). 
The first column gives the MAE, while the second column gives the mean percentage difference,
each including an average on the high-symmetry triangular route $O-$$\langle001\rangle$
$-O_{001}-$$\langle110\rangle$$-O_{111}-$$\langle111\rangle$$-O$
and the full 3D real-space grid. The average absolute AFQMC statistical
error, in value and in percentage, are listed at the first row as
a reference.}

\begin{tabular}{c|cc|cc}
\multirow{2}{*}{} & \multicolumn{2}{c|}{MAE ($\times10^{-4}$ $\mathrm{Ry}^{-3}$)} & \multicolumn{2}{c}{Mean percentage error (\%) }\tabularnewline
\cline{2-5} 
 & Triangular route & Full 3D grid & Triangular route & Full 3D grid\tabularnewline
\hline 
\textbf{AFQMC}\footnotemark[1]  & \textbf{1.04} & \textbf{0.76} & \textbf{0.29} & \textbf{0.30}\tabularnewline
LDA & 10.4 & 4.35 & 3.10 & 2.03\tabularnewline
PBE & 6.9 & 2.87 & 1.86 & 0.97\tabularnewline
PBEsol & 10.6 & 4.21 & 3.12 & 1.77\tabularnewline
B3LYP & 7.1 & 3.86 & 2.76 & 2.17\tabularnewline
PBE0 & 7.8 & 2.97 & 2.32 & 1.23\tabularnewline
\end{tabular}

\footnotetext[1]{AFQMC statistical error}
\end{table*}

\begin{table*}[b]
\caption{\label{tab:NaCl-dens-benchmark} Same as Table \ref{tab:Si-dens-benchmark}
but for NaCl.}

\begin{tabular}{c|cc|cc}
\multirow{2}{*}{} & \multicolumn{2}{c|}{MAE ($\times10^{-4}$ $\mathrm{Ry}^{-3}$)} & \multicolumn{2}{c}{Mean percentage error (\%) }\tabularnewline
\cline{2-5} 
 & Triangular route & Full 3D grid & Triangular route & Full 3D grid\tabularnewline
\hline 
\textbf{AFQMC}\footnotemark[1]  & \textbf{1.74} & \textbf{0.75} & \textbf{0.26} & \textbf{0.37}\tabularnewline
LDA & 29.0 & 6.64 & 4.47 & 5.13\tabularnewline
PBE & 13.7 & 1.43 & 0.86 & 0.81\tabularnewline
PBEsol & 19.0 & 3.11 & 2.40 & 2.49\tabularnewline
B3LYP & 12.9 & 1.27 & 0.67 & 0.61\tabularnewline
PBE0 & 11.7 & 2.10 & 1.62 & 1.30\tabularnewline
\end{tabular}

\footnotetext[1]{AFQMC statistical error}
\end{table*}

\begin{table*}[b]
\caption{\label{tab:Cu-dens-benchmark} Same as Table \ref{tab:Si-dens-benchmark}
but for fcc-Cu.}

\begin{tabular}{c|cc|cc}
\multirow{2}{*}{} & \multicolumn{2}{c|}{MAE ($\times10^{-4}$ $\mathrm{Ry}^{-3}$)} & \multicolumn{2}{c}{Mean percentage error (\%) }\tabularnewline
\cline{2-5} 
 & Triangular route & Full 3D grid & Triangular route & Full 3D grid\tabularnewline
\hline 
\textbf{AFQMC}\footnotemark[1]  & \textbf{9.5} & \textbf{4.3} & \textbf{0.36} & \textbf{0.46}\tabularnewline
LDA & 91.2 & 16.1 & 1.20 & 1.52\tabularnewline
PBE & 60.8 & 12.9 & 1.24 & 1.14\tabularnewline
PBEsol & 73.7 & 12.3 & 1.43 & 1.39\tabularnewline
B3LYP & 44.0 & 21.7 & 2.57 & 2.81\tabularnewline
PBE0 & 53.7 & 22.1 & 1.40 & 2.05\tabularnewline
\end{tabular}

\footnotetext[1]{AFQMC statistical error}
\end{table*}

\bibliographystyle{apsrev4-1}
\bibliography{Density_Paper_ver1.3}

\end{document}